\documentclass[aps,twocolumn,superscriptaddress]{revtex4-1}
\usepackage{graphicx}
\usepackage{longtable,array}
\usepackage{amsmath}
\usepackage{color}
\usepackage{float}
\usepackage{hyperref}
\usepackage{wasysym}
\usepackage{url}
\usepackage{dcolumn}% Align table columns on decimal point
\usepackage{bm}% bold math
\usepackage{amssymb}

\usepackage[T1]{fontenc} %added for Polish characters

\draft
\begin{document}

\title{Accounting for the detection response brings cavity ring-down spectroscopy to sub-permille accuracy
%Short title: Sub-permille cavity ring-down spectroscopy
}

\author{Rajesh Chell}
\affiliation{Institute of Physics, Faculty of Physics, Astronomy and Informatics, Nicolaus Copernicus University in Toru\'n, Grudziadzka 5, 87-100 Torun, Poland}
\author{Marcin Gibas}
\affiliation{Institute of Physics, Faculty of Physics, Astronomy and Informatics, Nicolaus Copernicus University in Toru\'n, Grudziadzka 5, 87-100 Torun, Poland}
\affiliation{Central Office of Measures, Physical Chemistry and Environment Department, Świętokrzyskie Laboratory Campus of the Central Office of Measures, Wrzosowa 46, 25-211 Kielce, Poland}
\author{Szymon W\'{o}jtewicz}
\affiliation{Institute of Physics, Faculty of Physics, Astronomy and Informatics, Nicolaus Copernicus University in Toru\'n, Grudziadzka 5, 87-100 Torun, Poland}
\author{Jakub Łasocha}
\affiliation{Institute of Physics, Faculty of Physics, Astronomy and Informatics, Nicolaus Copernicus University in Toru\'n, Grudziadzka 5, 87-100 Torun, Poland}
\author{Daniel Lisak}
\affiliation{Institute of Physics, Faculty of Physics, Astronomy and Informatics, Nicolaus Copernicus University in Toru\'n, Grudziadzka 5, 87-100 Torun, Poland}
\author{Agata Cygan}
\email{agata@fizyka.umk.pl}
\affiliation{Institute of Physics, Faculty of Physics, Astronomy and Informatics, Nicolaus Copernicus University in Toru\'n, Grudziadzka 5, 87-100 Torun, Poland}

\date{\today}

\begin{abstract}
For nearly four decades, cavity ring-down spectroscopy (CRDS) has been among the simplest, most widely used absorption techniques, requiring no calibration because it measures decay times, not intensities. Yet as accuracy requirements have tightened, CRDS has proven strikingly inconsistent: reported errors range from several percent to the permille level---the signature of an uncontrolled systematic effect so far worked around rather than resolved. This has confined applications demanding the highest accuracy---from testing \textit{ab initio} theory to atmospheric sensing and quantum-based metrology---to a handful of laboratories relying on advanced calibration or architectures, at the cost of simplicity. Here we show the main source was hiding in plain sight: the detection chain's transient response. It distorts every ring-down, biasing the retrieved decay times differently from instrument to instrument---a bias that routine truncation of the early decay masks rather than removes. We measure this response with a calibration-free beat-note method and build it into the ring-down model. Validation uses a well-characterized carbon monoxide transition and an independent frequency-based reference. Across four detection configurations, line-intensity biases of up to 14\% shrink 41- to 87-fold, the results converge, reaching up to sub-permille agreement with the reference. This holds down to 4~$\mu$s ring-down times, extending accurate CRDS to stronger absorption. Independent of the absorbing species and requiring no change to the CRDS setup, the approach turns a technique valued for its simplicity into one equally valued for its accuracy. Benchmark-quality measurement thus comes within reach of any CRDS laboratory.

\end{abstract}

\maketitle

\section*{Introduction}
\label{sec1}

\begin{figure*}[t]
\centering
\includegraphics[width=\textwidth]{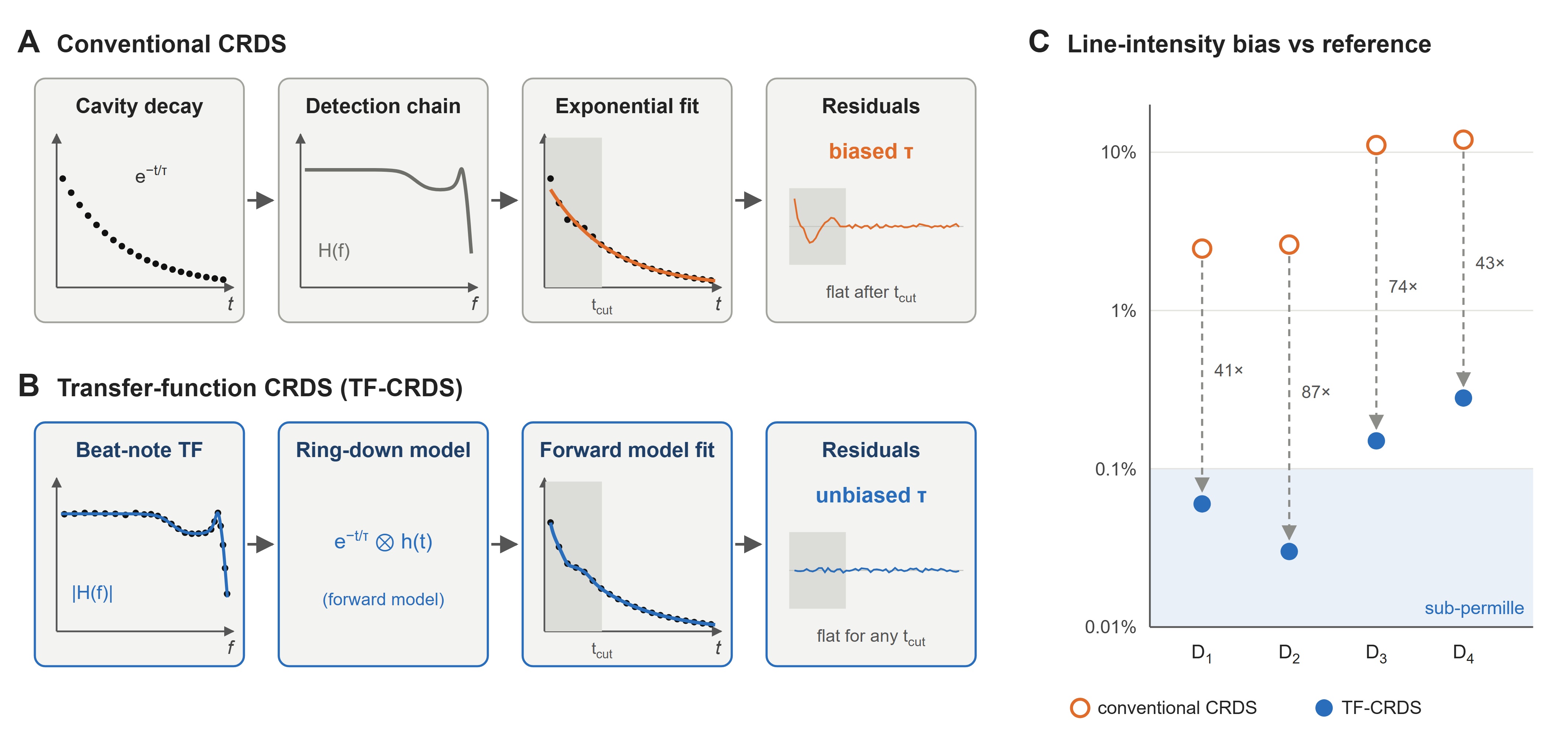}
\caption{\textbf{Concept of transfer-function CRDS.} (\textbf{A}) In conventional CRDS, the cavity decay is recorded through the detection chain, whose transfer function $H(f)$ may depart from unity at frequencies spanned by the ring-down. In that case, the recorded transient is a slightly distorted version of the cavity decay. An exponential function (orange) is fitted to this detection-modified transient after truncation of its early part ($t<t_{\mathrm{cut}}$, shaded). The fit leaves flat residuals after $t_{\mathrm{cut}}$, but extended into the truncated part it departs from the recorded points. Truncation thus hides the distortion from the residuals without removing it: the retrieved decay time is biased with and without truncation. (\textbf{B}) In TF-CRDS, $H(f)$ is measured with a calibration-free beat-note method and built into the ring-down model, in which the cavity decay is convolved with the impulse response $h(t)$ of the detection chain. This forward model (blue) reproduces the same recorded transient over its full length, including the part truncated in (A), leaves flat residuals for any truncation time, and returns the decay time of the cavity itself. (\textbf{C}) Measured line-intensity bias for the four detection configurations tested ($D_1$--$D_4$), averaged over truncation times, before (open symbols) and after (filled symbols) including the transfer function; the shaded band marks the sub-permille level. The bias is determined relative to an independent, frequency-based reference that is largely insensitive to detection-system distortions. Panels (A) and (B) are schematic: black points show the cavity decay, the recorded transient, and the measured transfer function, whereas orange and blue lines show the fits of conventional CRDS and TF-CRDS, respectively, together with their residuals.}
\label{Fig_concept}
\end{figure*}

Cavity ring-down spectroscopy (CRDS)~\cite{OKeefe1988,Romanini1997} is, in principle, everything a spectroscopic technique should be. For nearly four decades it has combined exceptional sensitivity with a simple, self-referenced measurement scheme that requires no external calibration standard. The absorption coefficient is derived directly from cavity decay times rather than from absolute detected light intensity and absorption path length. These properties have made CRDS one of the most widely used absorption techniques worldwide, applied across a broad range of molecular species and spectral regions. Reaching benchmark-level accuracy, however, has so far required exactly what CRDS avoids---advanced external calibration or specialized experimental architectures~\cite{Fleisher2019,Cygan2019,Bielska2022,CCQMP229}. As accuracy requirements have tightened, CRDS results have proven inconsistent across studies. Some report errors of a few percent~\cite{Wojtewicz2011,Fleisher2019,Cygan2025}, while others report permille-level agreement with independent reference methods~\cite{Huang2024}. This inconsistency points to an uncontrolled systematic effect, one large enough to be reflected in the scatter among values compiled in international reference databases such as HITRAN~\cite{HITRAN2017}, and one that has so far been worked around rather than resolved. As a result, conventional CRDS has limited applicability in the most demanding, ultra-high-accuracy measurements---precisely the applications for which its sensitivity and simplicity would otherwise make it the obvious choice.

Accurate measurements of molecular line intensities and line shapes underpin radiative transfer models used in Earth atmospheric monitoring~\cite{HITRAN2024} and in the interpretation of planetary, exoplanetary, stellar, and interstellar spectra~\cite{Tennyson2012,Canocchi2024}. Fundamental studies, including searches for variations in fundamental constants in distant astrophysical objects~\cite{Bagdonaite2015,Ubachs2018,Webb2011} and tests of quantum electrodynamics in simple molecular systems~\cite{Pachucki2025,Stankiewicz2026}, depend critically on this accuracy. So does precision gas metrology, including water amount-fraction measurements~\cite{Hashiguchi2023,Castrillo2024} and optical thermometry~\cite{Gotti2018,Lisak2025,Li2025}. Biases in integrated line area, line position, or line width propagate directly into systematic errors in the derived spectroscopic and thermodynamic quantities.

Atmospheric remote sensing places particularly stringent demands on spectroscopic reference data, requiring sub-permille accuracy in the parameters of key atmospheric species to ensure reliable satellite-based retrievals~\cite{Basu2011,Wunch2011,ODell2018}. Residual spectroscopic errors otherwise bias inferred atmospheric sources and sinks, distort long-term composition trends, and compromise assessments of the Earth's energy balance and climate projections.

High-fidelity spectroscopy is equally essential for validating \textit{ab initio} calculations. Accurate potential energy and dipole moment surfaces allow prediction of line positions, intensities, and line shapes~\cite{Zak2017,Polyansky2018,Kowzan2020,Sanchez2021}. Comparison with experiment~\cite{Hartmann2013,Birk2017,Rubin2022,Cygan2025,Bielska2022,CCQMP229} then provides a stringent test of quantum-mechanical models and guides their refinement, particularly for conditions that are difficult to realize experimentally, including high-temperature exoplanetary atmospheres, cold interstellar environments, and high-pressure systems~\cite{Tennyson2012}. Recent \textit{ab initio} calculations of line intensities at sub-permille accuracy~\cite{Bielska2022,CCQMP229} have extended this role into precision gas metrology. They have enabled next-generation optical standards for temperature, pressure, and gas concentration~\cite{Lisak2025,Li2025}, and absolute determinations of isotopic ratios~\cite{Fleisher2021}. These approaches exploit intrinsic molecular properties as SI-traceable observables, reducing reliance on conventional calibration standards and pointing toward quantum-defined spectroscopic metrology.

To date, however, permille- to sub-permille-level accuracy has been demonstrated only in a limited number of cases, providing critical reference data for testing theoretical models and validating spectroscopic parameters~\cite{Polyansky2015,Huang2024,Bielska2022,CCQMP229,Cygan2025,Adkins2025}. Extending this level of performance to a technique that is already simple, sensitive, and deployed in laboratories worldwide---without compromising the properties that make it valuable---would do more than add one more accurate method to the list. It would make benchmark-quality measurements routinely accessible outside a small number of specialized laboratories, and would let theory be tested far more widely than it currently can be.

This has not, so far, been achieved by resolving the underlying limitation directly. Its main source is neither the molecular absorption process itself nor well-known experimental artifacts such as photoreceiver saturation, cavity-mode interference, or incomplete laser extinction, which are routinely controlled, but systematic distortions introduced by the CRDS detection system. Finite detection bandwidth, detector and digitizer nonlinearities, and other electronic effects distort the recorded ring-down transient and bias the retrieved absorption parameters~\cite{Wojtewicz2011,Fleisher2019,Cygan2025}. Two previous approaches have addressed parts of this problem, each at a cost. Ref.~\cite{Fleisher2019} calibrated the digitizer against a highly linear reference digitizer from electrical metrology, enabling sub-permille-accuracy validation of \textit{ab initio} line intensities~\cite{Bielska2022,CCQMP229}. This approach, however, sacrifices the calibration-free character of CRDS, and such reference equipment is in practice available only at leading national metrology institutes. It was, moreover, demonstrated only over a limited range of relatively long ring-down times, well above those investigated here. More recently, heterodyne CRDS~\cite{Cygan2025} experimentally determined the transfer function of the full detection chain and used it to identify a regime of approximately flat detection response, thereby avoiding detection-induced distortions rather than modeling them. It enabled calibration-free, sub-permille measurements of CO and HD transitions in agreement with state-of-the-art \textit{ab initio} calculations, but requires an advanced laser-to-cavity lock and a substantially more complex architecture than conventional CRDS. In neither case was the measured detection response built into the model of a conventional ring-down measurement.

Here we show that this is the missing step. The transient response of the detection chain distorts every recorded ring-down and biases the retrieved decay times in an instrument-specific way. We find this response to be the dominant contribution to the systematic error budget once other, already well-characterized instrumental effects are accounted for. Because the distortion is most visible in the early part of the transient, routine truncation of that part---commonly used to remove switching artifacts---masks the distortion rather than removing it, and statistically flat residuals can coexist with biased time constants (Fig.~\ref{Fig_concept}A). We measure the transfer function (TF) of the complete detection chain, including the detector, electronics, and digitization stages, with a calibration-free beat-note method and build it directly into the ring-down model, an approach we refer to as transfer-function CRDS (TF-CRDS). Recorded transients are then described to the statistical-noise limit over their full length, with flat residuals for any truncation time. More importantly, the retrieved decay time is that of the cavity itself rather than an apparent one biased by the detection chain (Fig.~\ref{Fig_concept}B). Neither external calibration standards nor any change to the CRDS architecture is required.

To validate TF-CRDS, we apply it to a well-characterized CO absorption line measured with four CRDS detection configurations. As the reference, we use cavity mode dispersion spectroscopy (CMDS)~\cite{Cygan2015,Cygan2016,Cygan2019}, a frequency-based technique that is largely insensitive to detection-system distortions and has been validated at the sub-permille level against state-of-the-art \textit{ab initio} line-intensity calculations~\cite{Bielska2022,CCQMP229}. In TF-CRDS, line-intensity biases of up to 14\% shrink 41- to 87-fold, the agreement with CMDS is about 0.1\% on average across the four configurations and reaches up to the sub-permille level (Fig.~\ref{Fig_concept}C). These results hold down to ring-down times as short as 4~$\mu$s, extending accurate CRDS to stronger absorption---that is, to stronger lines and higher sample pressures, easing the metrological demands of low-pressure operation. Because the approach targets the detection hardware rather than the absorbing species, it is expected to apply to CRDS measurements of any molecule, and because it requires no external reference or change to the CRDS setup, it can be adopted by the many CRDS instruments already in use worldwide. These results identify the detection transfer function as the missing element in the quantitative description of CRDS. They turn a technique long valued for its simplicity into one that can be valued equally for its accuracy---extending the reach of benchmark-quality spectroscopy from a handful of specialized groups to any laboratory able to build a cavity ring-down setup.

\section*{Results}

\subsection*{Calibration-free detection transfer function}
\label{sec2}

\begin{figure}[t]
\centering
\includegraphics[width=0.47\textwidth]{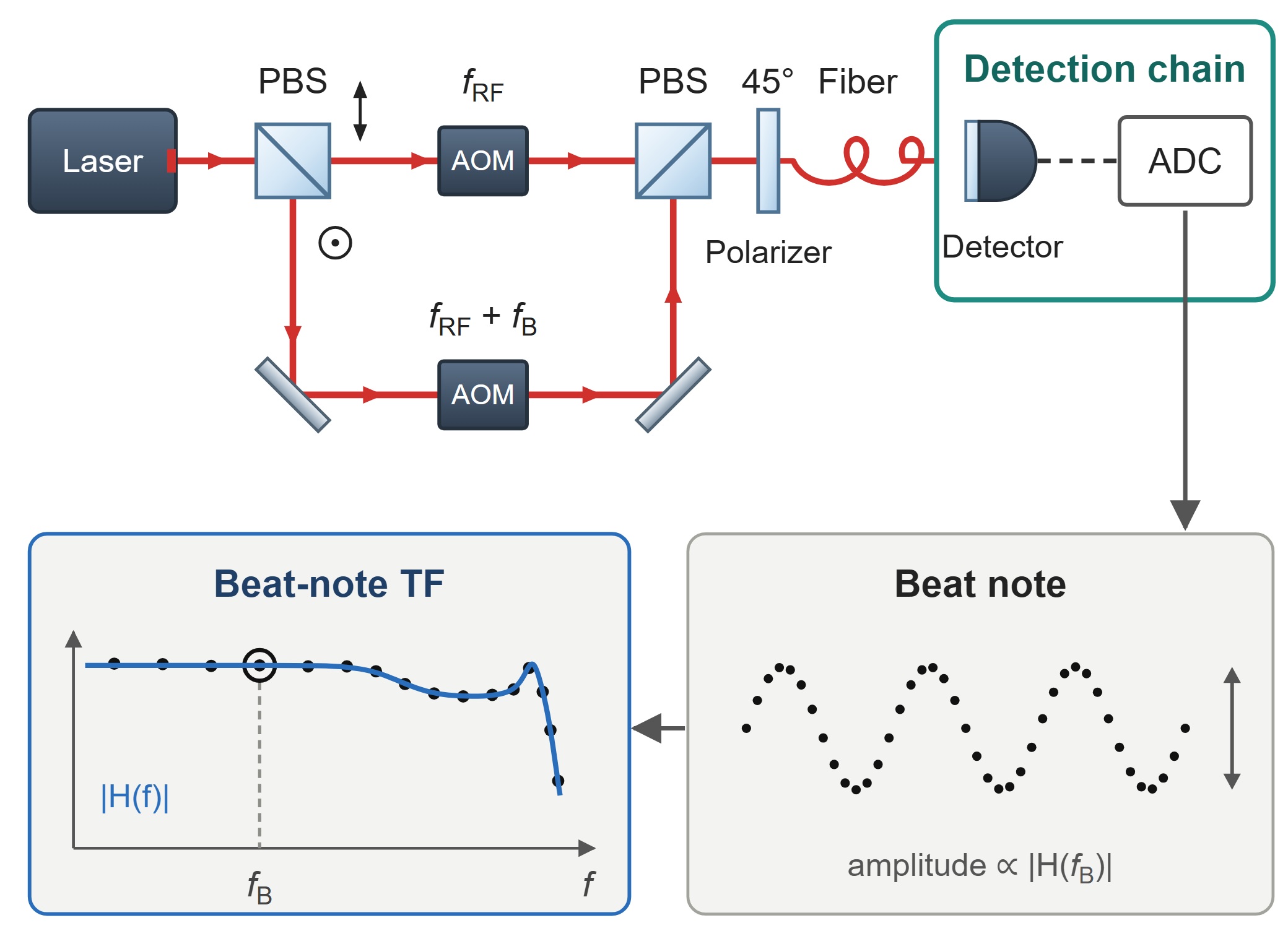}
\caption{\textbf{Calibration-free measurement of the detection transfer function.} A laser beam is split by a polarizing beam splitter into two orthogonally polarized beams, which are frequency shifted by AOMs driven at $f_{\mathrm{RF}}$ and $f_{\mathrm{RF}}+f_B$. The beams are recombined on a second PBS, projected onto a common polarization by a polarizer at 45$^{\circ}$, and coupled into a fiber to ensure optimal interference conditions. The resulting beat note at $f_B$ is recorded by the complete detection chain, and its amplitude is proportional to $|H(f_B)|$. Repeating the measurement for different values of $f_B$ yields the transfer function (TF) without any external calibration standard (black points, measured values; blue line, model). PBS, polarizing beam splitter; AOM, acousto-optic modulator; ADC, analog-to-digital converter.}\label{Fig_TF-setup}
\end{figure}

The missing element identified above---the frequency-dependent response of the complete detection chain---can be measured directly, without external calibration. We employ a well-established frequency-response approach from electronics, in which the system response is measured using sinusoidal signals of varying frequency. In the optical domain, these signals are generated by interfering two laser beams with an adjustable frequency offset. The experimental setup is shown in Fig.~\ref{Fig_TF-setup}. Two orthogonally polarized beams derived from the same laser are frequency-shifted by acousto-optic modulators driven at $f_{\mathrm{RF}}$ and $f_{\mathrm{RF}}+f_B$, recombined on a second polarizing beam splitter, projected onto a common polarization state by a polarizer, and delivered to the detector through a single-mode fiber. The resulting interference signal is $I=I_1+I_2+2\sqrt{I_1I_2}\cos{(2\pi f_Bt+\phi)}$, where $I_1$ and $I_2$ denote the beam intensities, and the last term represents the beat note at frequency $f_B$ with relative phase $\phi$. The measured beat-note amplitude, \(B_{12}^{m}(f_B)\), is equal to the product of the interference amplitude, \(2\sqrt{I_1I_2}\), and the magnitude of the transfer function of the complete detection chain at $f_B$, \(|H(f_B)|\). Assuming a linear detector amplitude response, the magnitude of the detection transfer function is obtained as

\begin{equation}
\label{eq1}
|H(f_B)|=\frac{B_{12}^m(f_B)}{2\sqrt{I_1^mI_2^m}},
\end{equation}
where the individual beam intensities \(I_1^m\) and \(I_2^m\) are measured by temporarily blocking the other beam. Importantly, no external calibration is required to determine the detection transfer function using this approach.

\begin{figure}[t]
\centering
\includegraphics[width=0.46\textwidth]{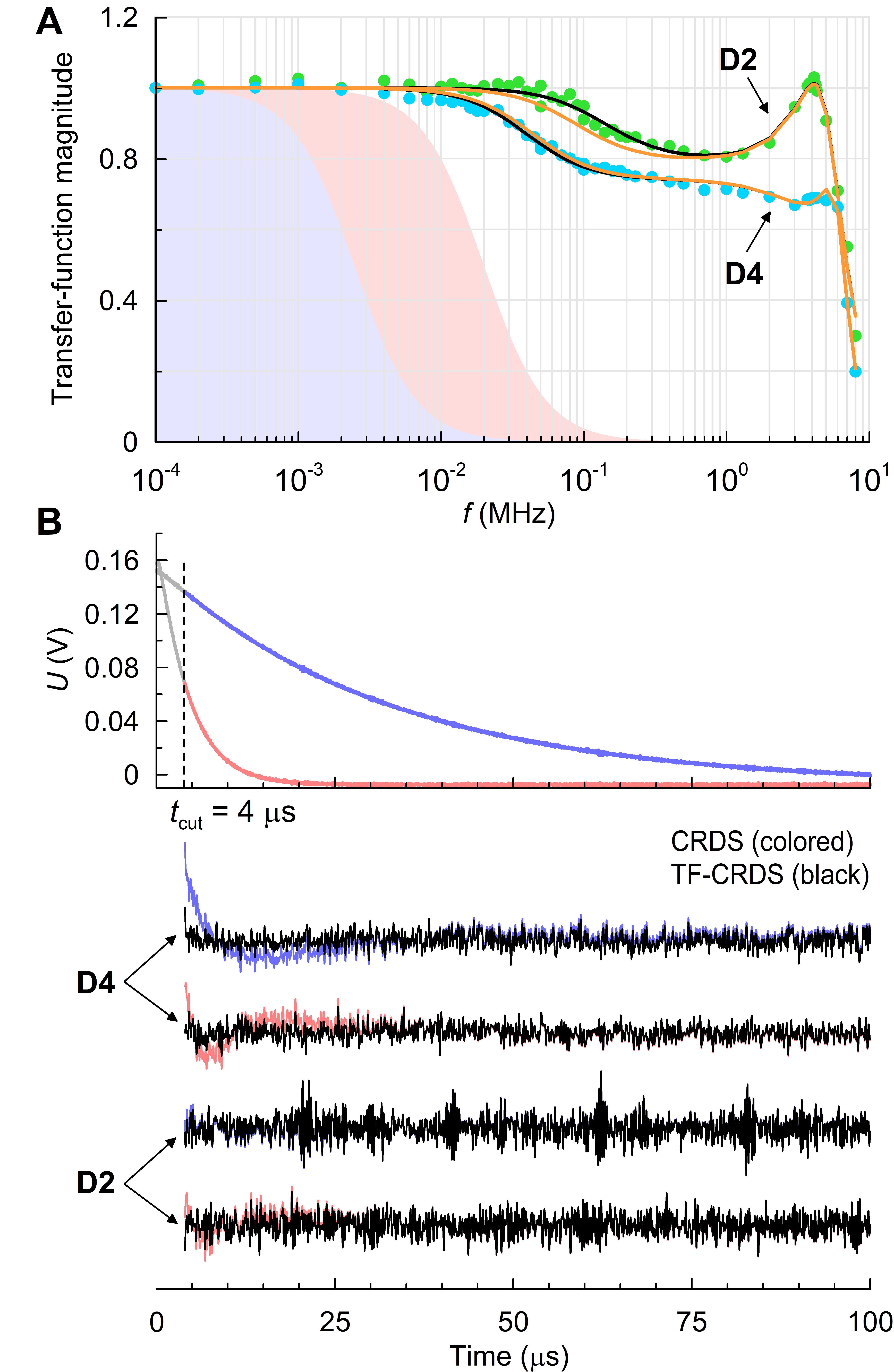}
\caption{\textbf{Detection transfer functions and their effect on ring-down residuals.} (\textbf{A}) Experimental transfer functions (TF) for two detection configurations, $D_2$ and $D_4$ (dots), with $|H(f)|$ model from Eq.~\ref{eq2} fitted to the experimental TF (black) and refined by multi-ring-down fitting (orange). Shaded regions are Fourier spectra of the ring-down signals shown in (B), color-coded accordingly. (\textbf{B}) Ring-down signals averaged over $100$ transients for $D_2$ and $D_4$ with time constants $32$ and $4~\mu\mathrm{s}$ (blue and red), and residuals from exponential CRDS fits (same colors) and TF-CRDS (black) fits using $S_{\mathrm{fit}}(t)$ with refined $H(f)$. Ring-downs are truncated at $t_{\mathrm{cut}}=4~\mu\mathrm{s}$. TF-CRDS residuals RMS: $0.11$--$0.15~\mathrm{mV}$.} \label{Fig_TF_plot_v2_selected2}
\end{figure}

Examples of experimentally determined transfer functions for two CRDS detection configurations (\(D_2\) and \(D_4\); see the ``CRDS and CMDS spectroscopy setup'' section) are shown in Fig.~\ref{Fig_TF_plot_v2_selected2}A. In both cases, the response remains close to unity up to approximately \(10\)~kHz, then gradually deviates from unity at higher frequencies, and drops steeply above approximately \(6\)~MHz, the upper limit of the detection bandwidth. The measured transfer functions are well described by the model of Eq.~\ref{eq2}, introduced in the ``Transfer-function measurement and model'' section (black curves in Fig.~\ref{Fig_TF_plot_v2_selected2}A). This model combines two low-pass components: a first-order component, which describes the gradual decrease of the response well below the detection bandwidth (here with cutoff frequencies \(f_c\) of 37--83~kHz), and a higher-order resonant component describing the fast part of the response up to the bandwidth limit. Because the ring-down spectra extend only up to a few hundred kilohertz (shaded areas in Fig.~\ref{Fig_TF_plot_v2_selected2}A), the ring-down signals are affected mainly by the first-order component rather than by the detection bandwidth. In the time domain, this first-order component corresponds to a slow part of the detector response. Depending on the detection configuration (\(D_1\)--\(D_4\)), this slow part carries 15--28\% of the signal and relaxes with a time constant \(1/(2\pi f_c)\approx 2\)--\(4~\mu\mathrm{s}\), comparable to the shortest ring-down times measured here (see the ``Fast and slow components of the detection response'' section). Further details of the transfer-function measurement and modeling, including the treatment of static nonlinearities, are given in Materials and Methods.

\subsection*{Transfer-function CRDS}
\label{sec3}

Once measured, the detection transfer function is built directly into the ring-down fit model. The exponential decay function, $S_{\mathrm{ed}}(t)$, is Fourier transformed, multiplied by the transfer function $H(f)$, and transformed back into the time domain according to $S_{\mathrm{fit}}(t)=\mathcal{F}^{-1}\!\left\{\mathcal{F}\!\left[S_{\mathrm{ed}}(t)\right]H(f)\right\}$, with further details, including a comparison with deconvolution, provided in the ``Forward-model implementation'' section. This forward-modeling approach is equivalent to convolving the exponential decay with the impulse response $h(t)$ of the detection chain in the time domain (Fig.~\ref{Fig_concept}B). Reconvolution with a measured instrument response is routine in time-resolved fluorescence spectroscopy~\cite{Lakowicz2006}. In CRDS, by contrast, the detection response has commonly been assumed to distort only the earliest part of the transient, which is then truncated (Fig.~\ref{Fig_concept}A). Where it has been modeled, it has been described by a single time constant set by the detector bandwidth~\cite{Gong2007}, which matters only when it approaches the ring-down time. Here, this condition is met not by the bandwidth-limited response but by the slow part of the detector response. Even bi-exponential models that include such a slow part capture its effect only approximately and cannot guarantee sub-permille accuracy (see the ``Fast and slow components of the detection response'' section). TF-CRDS therefore fits the recorded ring-downs with \(S_{\mathrm{fit}}(t)\), which contains the complete measured transfer function.

An important feature of TF-CRDS is that the parameters of $H(f)$ (here, the cutoff frequency $f_c$) can be further refined by jointly fitting multiple ring-down transients with a common transfer-function model (see the ``Multi-ring-down refinement'' section). This refinement accommodates small changes in the shape of the transfer function caused by beam-alignment fluctuations at the detector and by level-dependent static nonlinearities. Both effects modify the transfer function obtained from the beat-note measurement in ways that are difficult to distinguish from the intrinsic frequency-dependent response of the detection chain. The refined transfer functions agree closely with those fitted directly to the measured data (compare black and orange curves in Fig.~\ref{Fig_TF_plot_v2_selected2}A), but even such small differences matter when ring-down time constants are determined at permille-level accuracy. Level-dependent static nonlinearities are themselves small: the shape of the measured transfer function remained unchanged when the laser power was reduced tenfold, showing that it does not depend on the signal level. Independent evidence that frequency-dependent detection effects dominate over static nonlinearities comes from recent heterodyne CRDS measurements \cite{Cygan2025}, in which shifting the detection to the flat-response region of the detector bandwidth reduced the \(\sim\)1\% bias of conventional CRDS to the sub-permille level and brought the results into agreement with \textit{ab initio} line intensities.

Figure~\ref{Fig_TF_plot_v2_selected2}B compares ring-down fits and corresponding residuals obtained using the conventional exponential decay model (CRDS) and the TF-CRDS model $S_{\mathrm{fit}}(t)$ with refined $H(f)$. In CRDS, residual structures (colored) emerge when the Fourier spectrum of the ring-down signal extends into frequency regions where \(H(f)\) deviates from unity (shaded areas in Fig.~\ref{Fig_TF_plot_v2_selected2}A). Such distortions become more pronounced for short ring-down times due to their broader Fourier spectra. In TF-CRDS, these distortions are reproduced by the model, yielding flat residuals (black) consistent with statistical noise.

\subsection*{Validation of TF-CRDS accuracy}
\label{sec4}

Statistically flat ring-down residuals are not sufficient to validate the accuracy of TF-CRDS, as subtle detector nonlinearities, such as power-law-type responses, may bias the retrieved ring-down time constants without leaving any trace in the residuals \cite{Fleisher2019,Cygan2021}. We therefore validate TF-CRDS using the well-characterized CO (3--0)-band \(R(23)\) transition. Measurements were performed with four detection configurations ($D_1$--$D_4$) using both CRDS and CMDS in the linear (unsaturated) absorption regime, covering ring-down times from 32~\(\mu\mathrm{s}\) down to 4~\(\mu\mathrm{s}\). The frequency-based CMDS technique \cite{Cygan2015,Cygan2016,Cygan2019}, benchmarked against \textit{ab initio} calculations for this transition \cite{Bielska2022,CCQMP229,Cygan2025}, provides an independent reference for validating the accuracy of TF-CRDS. For each detection configuration, CRDS and CMDS measurements were performed simultaneously under identical experimental conditions, providing a CMDS reference for every CRDS measurement. Because, for linear absorption, the integrated line areas obtained with both techniques are proportional to the product of line intensity and absorber number density, and the latter is common to the two simultaneous measurements, their ratio directly reflects the ratio of line intensities, largely independent of pressure and temperature uncertainty. Here, TF-CRDS time constants are retrieved using the \(S_{\mathrm{fit}}(t)\) model with the refined \(H(f)\). Detailed experimental conditions and measurement procedures are provided in the ``CRDS and CMDS spectroscopy setup'' section.

\begin{figure}[t]
\centering
\includegraphics[width=0.44\textwidth]{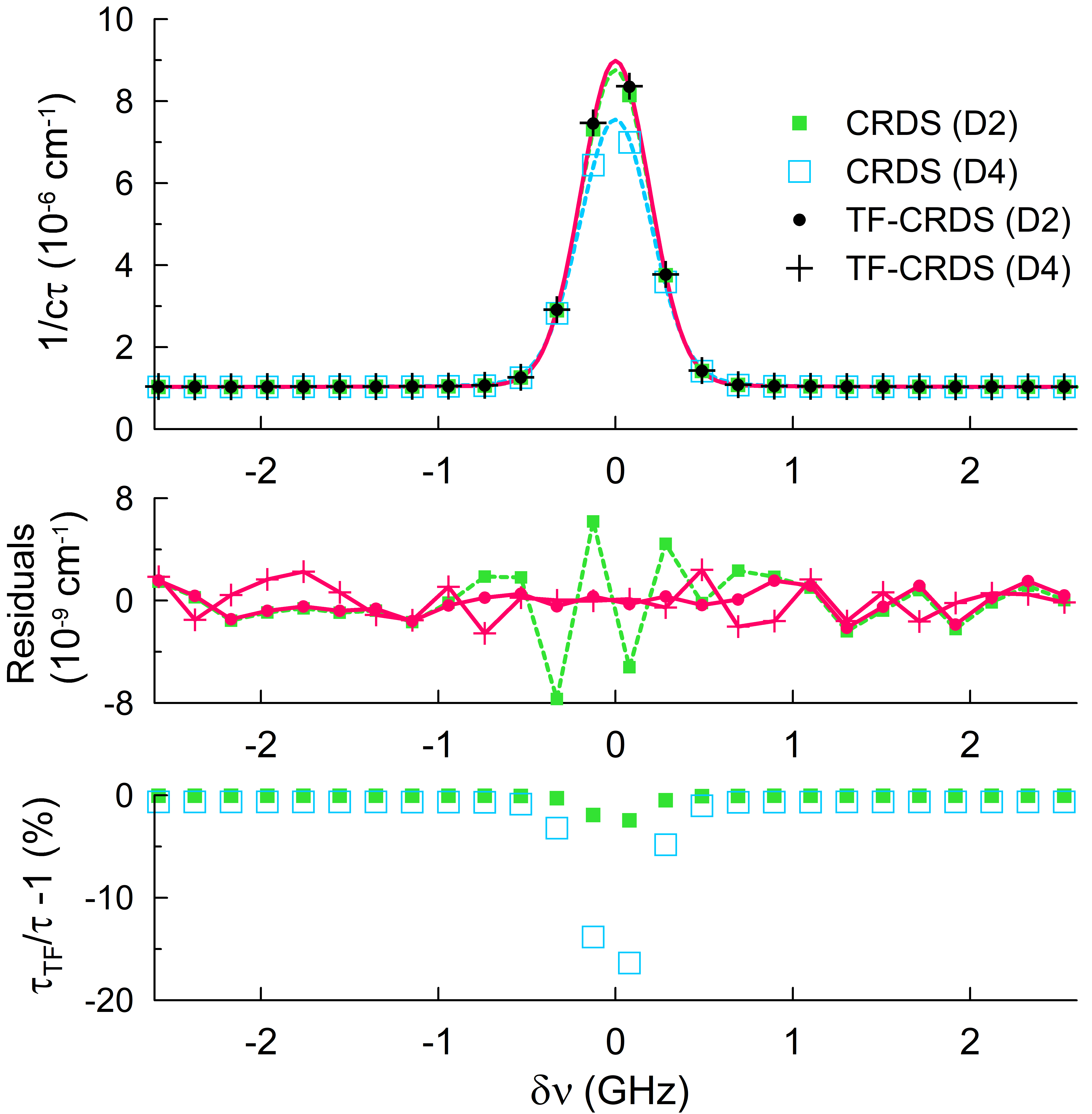}
\caption{\textbf{CO absorption spectra obtained with conventional CRDS and with TF-CRDS.} CRDS spectra of CO (3--0)-band \(R(23)\) transition recorded at a pressure of 0.67~kPa for detection configurations $D_2$ and $D_4$ (colored \scalebox{0.6}{$\blacksquare$} and \scalebox{0.8}{$\square$}) with fits of the speed-dependent hard-collision profile (dashed lines), and TF-CRDS spectra (black \scalebox{0.8}{\(\bullet\)} and \scalebox{0.9}{$+$}) with corresponding fits (red solid lines). Residuals are shown below with the same colors and symbols; $D_4$ residuals are omitted for clarity. The lowest panel shows the relative difference between the ring-down time constants retrieved with TF-CRDS, \(\tau_{\mathrm{TF}}\), and with conventional CRDS, \(\tau\). Ring-downs are truncated at $t_{\mathrm{cut}}=4~\mu\mathrm{s}$.} \label{Fig_CO_line_plot_v2_selected}
\end{figure}

Figure~\ref{Fig_CO_line_plot_v2_selected} compares CRDS and TF-CRDS absorption spectra of the CO (3--0)-band \(R(23)\) line for two selected detection configurations (\(D_2\) and \(D_4\)), whose transfer functions are shown in Fig.~\ref{Fig_TF_plot_v2_selected2}A. As discussed above, ring-down transients are distorted when their Fourier spectra overlap with frequency regions where \(H(f)\) deviates from unity, attenuating the corresponding signal components. In CRDS, this distortion leads to an artificial increase of the retrieved ring-down time constant and, consequently, to an underestimation of the absorption. The effect is particularly pronounced for \(D_4\), where the deviation of \(H(f)\) from unity is largest, and becomes stronger for shorter ring-down times, resulting in the largest bias at the absorption-line center. Consequently, CRDS exhibits structured line-shape fit residuals, which disappear in TF-CRDS (for the line-shape analysis, see the ``CRDS and CMDS spectroscopy setup'' section). Comparison with TF-CRDS shows that the conventional ring-down time constants are overestimated: the TF-CRDS values are shorter by \(0.04\%\) at the baseline up to \(2.4\%\) at the line center for \(D_2\), and by \(0.6\%\) up to \(16\%\) for \(D_4\). Notably, the smaller TF distortion for \(D_2\), largely hidden by statistical noise in the residuals (see Fig.~\ref{Fig_TF_plot_v2_selected2}B), still produces a percent-level bias, demonstrating that statistically flat residuals do not guarantee unbiased CRDS results.

\subsection*{Sub-permille accuracy and extended dynamic range}

\begin{figure}[t]
\centering
\includegraphics[width=0.45\textwidth]{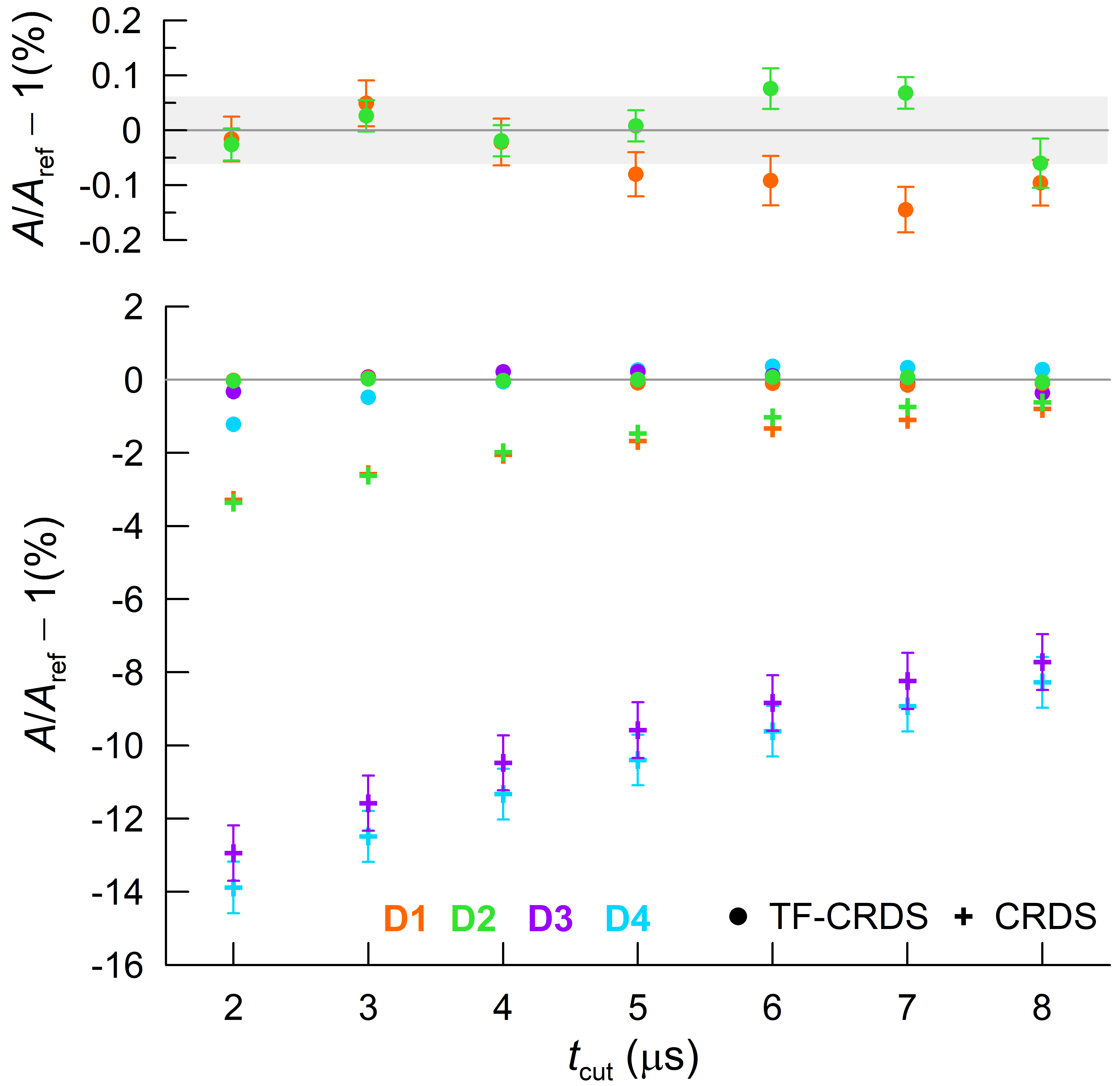}
\caption{\textbf{Line-intensity accuracy improvement and extended dynamic range enabled by the transfer function.} The integrated line areas of the CO (3--0)-band $R(23)$ line obtained from absorptive CRDS spectra, $A$, relative to the dispersive CMDS reference areas, $A_{\mathrm{ref}}$, for all detection configurations are shown as a function of the initial truncation time $t_{\mathrm{cut}}$ of the ring-down transient (colored \scalebox{0.9}{$+$}). TF-CRDS results are also shown (colored \scalebox{0.8}{\(\bullet\)}). The top panel shows the TF-CRDS results for $D_1$ and $D_2$ on an expanded vertical scale. The gray-shaded area indicates the spread of individual CMDS-based $A_{\mathrm{ref}}$ references.} \label{Fig_Accuracy_plot_v5_selected}
\end{figure}

The integrated areas of the CO (3--0)-band $R(23)$ line obtained from absorptive CRDS spectra, $A$, relative to the dispersive CMDS reference areas, $A_{\mathrm{ref}}$, for all detection configurations are shown in Fig.~\ref{Fig_Accuracy_plot_v5_selected} as a function of the initial truncation time $t_{\mathrm{cut}}$ of the ring-down transient. Transient truncation is commonly used to remove early-time distortions, such as those caused by delays between data acquisition and laser extinction, but may also mask systematic effects, including detection-induced distortions (Fig.~\ref{Fig_concept}A). These distortions persist beyond typical truncation times, because the slow part of the detector response relaxes on the same microsecond scale (see the ``Fast and slow components of the detection response'' section). The observed dependence of the \(A/A_{\mathrm{ref}}\) ratio on \(t_{\mathrm{cut}}\) reveals systematic distortions in CRDS (shown as $+$). The bias is largest at the shortest truncation times, reaching about \(-14\%\) for \(D_3\) and \(D_4\), and persists even at \(t_{\mathrm{cut}}=8~\mu\mathrm{s}\), where it still amounts to \(-0.7\%\) for \(D_1\) and \(D_2\) and \(-8\%\) for \(D_3\) and \(D_4\). In contrast, the TF-CRDS results (shown as \(\bullet\)) are essentially independent of \(t_{\mathrm{cut}}\), as expected for the present detection systems, where the distortion is dominated by the frequency-dependent detection response. The residual bias is strongly reduced over the investigated range of \(t_{\mathrm{cut}}\) for all configurations, remaining within about \(0.15\%\) for \(D_1\) and \(D_2\) at every truncation time (see top panel of Fig.~\ref{Fig_Accuracy_plot_v5_selected}). The \(D_3\) and \(D_4\) configurations, by contrast, provide deliberately challenging cases, exhibiting the largest transfer-function distortions. For \(D_4\), a small residual dependence on \(t_{\mathrm{cut}}\) remains at the shortest truncation times, about 1\% at \(2~\mu\mathrm{s}\) and 0.4\% at \(3~\mu\mathrm{s}\), and practically vanishes for longer truncation times. In this configuration, the transfer function departs from unity already below 10~kHz (Fig.~\ref{Fig_TF_plot_v2_selected2}A), and its slow component relaxes most slowly (\(\tau_c\approx4.3~\mu\mathrm{s}\)). At the earliest times, where the free response of the slow component is largest, potential small inaccuracies of the transfer-function model have the greatest effect. Averaged over the investigated \(t_{\mathrm{cut}}\) range, the residual bias of TF-CRDS is \(0.06\%\), \(0.03\%\), \(0.15\%\), and \(0.28\%\) for \(D_1\), \(D_2\), \(D_3\), and \(D_4\), respectively. Compared with the conventional CRDS values averaged over the same range, this corresponds to a 41-, 87-, 74-, and 43-fold reduction. The individual CMDS-based \(A_{\mathrm{ref}}\) references exhibit a spread below \(0.06\%\) (gray-shaded area in Fig.~\ref{Fig_Accuracy_plot_v5_selected}) under stable measurement conditions, providing a reliable basis for validating the accuracy of TF-CRDS at the sub-permille level.

These results are consistent with, and help resolve, the inconsistency noted in the Introduction. Recent work by Huang et al. \cite{Huang2024} demonstrated permille-level agreement between CRDS and CMDS, in contrast to earlier reports of percent-level discrepancies \cite{Wojtewicz2011,HITRAN2017,Fleisher2019,Cygan2025}. This agreement does not exclude detection-response distortions, which are expected to remain small for ring-down times down to approximately 22~$\mu$s---inferred from the absorption data of Huang et al. \cite{Huang2024}---but become significant for substantially shorter transients. TF-CRDS extends the dynamic range of CRDS, enabling sub-permille accuracy even for ring-down times as short as 4~$\mu$s. This removes a practical constraint that has traditionally favored long ring-down times: stronger lines and higher sample pressures become accessible, easing the metrological demands of low-pressure operation without sacrificing accuracy.

\section*{Discussion}
\label{sec5}

%The detection transfer function is the missing element in the quantitative description of CRDS. 
Our results point to the detection chain, rather than the absorption measurement itself, as the dominant source of the systematic scatter in reported CRDS accuracy, which ranges from several percent~\cite{Wojtewicz2011,Fleisher2019,Cygan2025} to the permille level~\cite{Huang2024}. The detection response biases the retrieved ring-down time constants, and hence the measured absorption, by an amount set by how far the Fourier spectrum of the ring-down extends into the frequency range where the response departs from unity. This bias therefore depends on the ring-down times, the photodetector, its gain setting, and the beam alignment---all of which differ from one laboratory to the next, and even between nominally identical instruments. In our measurements, two photodetectors of the same model, differing only in age and usage history, produced line-area biases that differed by an order of magnitude under otherwise identical conditions, and the detector with the lower noise produced the larger distortion. Neither the nominal bandwidth nor the noise performance of a detector predicts this bias; the measured transfer function does. Measurements made with long ring-down times and favorable detectors can thus reach permille-level agreement~\cite{Huang2024}, whereas the same technique applied under less favorable conditions yields percent-level errors---for reasons not readily apparent in a standard exponential fit of the ring-down.

Including the detection response in the ring-down model is therefore essential for two distinct purposes: obtaining undistorted ring-down transients and obtaining correct time constants. The two are not equivalent. For configuration $D_2$, the distortion appears in the residuals of the conventional exponential fit only in the first few microseconds of the transient. Truncating this part, a common remedy, makes the residuals look flat at the present signal-to-noise ratio, yet the bias remains: line-area biases of 0.7\% ($D_1$, $D_2$) and 8\% ($D_3$, $D_4$) persisted even after 8~$\mu$s of truncation. Truncation thus hides the distortion rather than removing it. A static nonlinearity of power-law type is a still more stringent case: it preserves the exponential form of every decay while rescaling its time constant, and therefore leaves no signature in the residuals at any truncation time or signal-to-noise ratio~\cite{Fleisher2019,Cygan2021}. The quality of a ring-down fit is therefore not a measure of its accuracy. Accuracy requires a model that reproduces the transient as it is actually recorded through the complete detection chain, including slow components that a single detector response time cannot describe, together with an independent test of the time constants that this model returns. Here, the transfer function provides the first, and the simultaneous CMDS reference---which relies on frequency rather than intensity and is therefore largely insensitive to the detection response---provides the second. Together they show that TF-CRDS retrieves the decay rate of the cavity itself, rather than an apparent decay rate shaped by the detector, with line areas that no longer depend on the choice of truncation time.

At sub-permille agreement with this reference, TF-CRDS, run on unmodified conventional CRDS hardware, performs on par with the techniques that have validated \textit{ab initio} line intensities with the highest accuracy to date~\cite{Bielska2022,CCQMP229}---cavity mode dispersion spectroscopy~\cite{Cygan2015,Cygan2019} and calibrated CRDS~\cite{Fleisher2019}---as well as heterodyne CRDS~\cite{Cygan2025}. This accuracy is maintained down to ring-down times of 4~$\mu$s, well below the shortest ring-down times of approximately 22~$\mu$s in the permille-level comparison of Ref.~\cite{Huang2024}. The range of accurate CRDS measurements thereby extends to stronger absorptions and higher pressures, easing the metrological demands of low-pressure operation, and to lower-finesse cavities, such as those used in spectral regions where high-reflectivity mirrors are less readily available.

Among CRDS-based approaches, the calibrated and heterodyne variants also reach sub-permille accuracy, but each at a cost. Calibrated CRDS characterizes the digitizer against a highly linear reference digitizer used in electrical metrology, using synthetic exponential decay signals from an arbitrary waveform generator~\cite{Fleisher2019}. It therefore requires electrical reference equipment in addition to the spectrometer. Heterodyne CRDS~\cite{Cygan2025} additionally provides simultaneous access to the dispersive part of the line shape and avoids detection distortions by shifting the measured signal into a flat region of the detection response. However, it requires a substantially more complex optical setup, including a demanding laser-to-cavity lock whose quality directly limits its performance. TF-CRDS needs neither. It works with the same simple hardware already found in conventional CRDS, requires no more demanding laser-to-cavity lock, and models the response of the complete detection chain rather than avoiding it. In exchange for giving up simultaneous dispersive information, it retains the sensitivity, favorable long-term noise performance, and experimental simplicity that have made CRDS one of the most widely used absorption techniques---now at benchmark accuracy as well.

Because it requires no calibration standard and no change to the CRDS architecture, TF-CRDS is, in principle, available to any laboratory already operating conventional CRDS, and could become a universal approach to quantitative CRDS rather than a specialized variant of it. This could substantially widen the experimental basis of precision molecular spectroscopy, with more laboratories independently testing \textit{ab initio} theory and fundamental physics. It could also directly support multi-laboratory measurement campaigns~\cite{Bielska2022,CCQMP229}, which were organized precisely because researchers were aware of percent-level disagreements between laboratories. Such campaigns need a method that is simple enough to be implemented in many laboratories and accurate enough to resolve sub-permille differences; TF-CRDS meets both requirements.

The same principle is not limited to ring-down decays. Any cavity transient recorded through a detection chain is subject to the same distortion, including the buildup of the intracavity field. This is directly relevant to cavity buildup dispersion spectroscopy~\cite{Cygan2021} and heterodyne CRDS~\cite{Cygan2025}, in which the cavity mode is retrieved from the Fourier spectrum of a beat signal recorded during the buildup or decay of the intracavity field. The dispersive spectrum, obtained from the center frequency of the mode, is less sensitive to the detection response, because it relies on a frequency measurement. The absorptive spectrum, however, is obtained from the width of this Fourier spectrum, which corresponds to the cavity-mode width and is set by the envelope of the beat signal; the less this envelope is distorted, the more accurately the mode width is recovered. Incorporating the transfer function into the models of these signals should therefore improve the accuracy of the absorption spectra obtained with both techniques and, in heterodyne CRDS, could relax the requirement that the beat signal lie within a flat region of the detection response.

The implications extend beyond spectroscopy itself. For atmospheric remote sensing, which requires spectroscopic parameters of key species at the sub-permille level~\cite{Basu2011,Wunch2011,ODell2018}, TF-CRDS offers a broader base of independently validated reference data. For metrology, it opens a simpler route toward the next generation of optical standards for temperature, pressure, and gas concentration~\cite{Gotti2018,Hashiguchi2023,Castrillo2024,Lisak2025,Li2025} and toward spectroscopic isotope-ratio measurements~\cite{Fleisher2021}, all of which depend on accurate line intensities and line shapes. For environmental and industrial monitoring, commercial CRDS analyzers, widely deployed for atmospheric and industrial trace-gas measurements, share the same underlying detection chain and could adopt TF-CRDS wherever their application calls for improved quantitative accuracy.

The validation reported here was designed so that the detection chain is the only variable under test. Because TF-CRDS targets the detection hardware rather than the absorbing species, its validity is not tied to the transition used here, provided that the absorption itself remains linear, as it does in the present measurements (see the ``CRDS and CMDS spectroscopy setup'' section). The four detection configurations were built from one detector model and one digitizer, a basis that implementations in other laboratories will broaden further. They nevertheless test the method under markedly different conditions: the line-intensity biases they cause in conventional CRDS differ by up to an order of magnitude.

Despite this spread, TF-CRDS reduced these biases by comparable factors, 41- to 87-fold. The residual bias of TF-CRDS relative to CMDS therefore follows that of conventional CRDS: it is sub-permille for $D_1$ and $D_2$ and 0.15\% and 0.28\% for $D_3$ and $D_4$, the deliberately challenging cases whose transfer functions depart most from unity. A flatter detection response still pays off, and only the measured transfer function reveals which detector provides it. Noise performance alone cannot reveal it: of the two nominally identical detectors, the lower-noise one produced the larger bias. How far the transfer function may depart from unity while TF-CRDS retains permille-level accuracy remains an open question. A stringent test would use detectors whose transfer function departs strongly from unity already at low frequencies, such as AC-coupled detectors, which do not transmit the slowest components of the ring-down at all. If TF-CRDS passes such tests, detectors could be chosen for low noise, speed, or cost rather than for a flat response---a prospect of particular interest for compact and commercial CRDS instruments.

The forward-modeling methodology itself does not depend on how the transfer function is acquired, provided that it is determined with sufficient accuracy. Here, it was measured with two independently driven acousto-optic modulators (AOMs), but the same beat-note measurement can be performed with a single AOM: either the free-space AOM already present in a conventional CRDS setup, followed by a single-mode fiber, or a fiber-coupled AOM. In both cases, the modulator is driven simultaneously at two radio frequencies separated by $f_B$, so that the beat note of the two-tone RF signal is transferred directly onto the optical beam. For the tightly focused beams typical of fast-switching AOMs, the slight change in diffraction angle with radio frequency affects the fiber coupling only weakly, and any residual change is captured by the beam powers, measured behind the fiber, that enter the normalization. A second, faster route uses the modulator exactly as it is used to initiate a ring-down: the detection response is obtained from a single switch-off of a beam that bypasses the cavity, recorded with the same detection settings as the ring-downs. Because one such transient covers the entire relevant frequency range at once, no scan of the beat frequency is needed. Both implementations are under development in our laboratory and will be reported separately. A technical limitation common to all implementations is that the transfer function depends on how the beam leaving the cavity is focused and aligned onto the detector's active area, consistent with the reported spatial inhomogeneity of the photodetector temporal response~\cite{Cao2019}. It must therefore be re-measured whenever the direction of the beam leaving the cavity changes. Because the measurement leaves the detection chain untouched---and, with the single-transient approach, takes little more time than a ring-down acquisition---it can readily be repeated as part of routine operation.

Two effects lie outside the present forward model, and both can be brought into it. The first concerns progressively shorter ring-down times. The model assumes that the AOM switches the beam off faster than the detection system can respond, so that the modulator itself need not be included in the effective detection response. The switch-off time of an AOM is set by the transit time of the acoustic wave across the optical beam, and hence by how tightly the beam is focused in the modulator crystal. Once this time becomes comparable to the response time of the detection system, which matters increasingly as ring-down times shorten, the modulator must be incorporated into the effective detection-system model---a constraint that does not arise in the steady-state operation of CMDS. The agreement with CMDS down to 4~$\mu$s indicates that the modulator, with a switch-off time of about 50~ns, was fast enough here to be neglected in the detection model, and tighter focusing in the modulator can maintain this down to still shorter ring-down times.

The second is a static nonlinearity of power-law type. Level-dependent static nonlinearities alter the shape of the transient differently at different signal levels and are therefore accounted for by the multi-ring-down refinement. A power-law response, by contrast, is scale-invariant: as noted above, it leaves every transient exponential and rescales all time constants by the same factor, independently of amplitude, ring-down time, and truncation, so that no linear transfer function can represent it. For $D_1$ and $D_2$, the sub-permille agreement with CMDS limits any such contribution in the present data to below $10^{-3}$. Where it must be quantified, this can again be done without external calibration and with the same two-beam arrangement used here for the transfer-function measurement. Because the powers of orthogonally polarized beams add exactly, comparing the detector signal for the two orthogonally polarized beams, recorded together and separately, reveals any departure from linearity, including a power-law response---the principle of the well-established superposition method for testing detector linearity~\cite{Kauten2014}. The forward model can then be extended by a power-law term, with the power-law exponent taken from this measurement, so that both the dynamic and the static response of the detection chain are described without any external standard.

Applying TF-CRDS in practice requires only a few steps. The transfer function is measured with the detector aligned as for the ring-down measurements and at comparable signal levels, and is then refined by the multi-ring-down fit using transients that span the ring-down and truncation times of the spectrum. Flat residuals and line areas independent of the truncation time indicate that frequency-dependent and level-dependent distortions have been removed, while a power-law component, which these checks cannot reveal, is tested with the same two beams using the superposition method. An independent reference---a frequency-based method, or a transition whose intensity is known at the sub-permille level from \textit{ab initio} calculations, such as the CO line used here~\cite{Bielska2022,CCQMP229,Cygan2025}---can then serve as additional validation rather than as a calibration. Reporting the measured transfer function alongside CRDS results would further allow the detection contribution to be assessed in multi-laboratory comparisons and give reference compilations such as HITRAN~\cite{HITRAN2024} a basis for judging the reliability of CRDS-derived parameters. For commercial analyzers, the transfer function could be measured at the factory, re-measured after servicing or realignment, and built into the instrument's data processing.

Conventional CRDS has been limited not by its principle but by an unmodeled property of its detection chain, which made its accuracy depend on the instrument rather than on the method. By measuring this property without external calibration and building it into the ring-down model, we reduced detection-induced biases by up to 87-fold and reached up to sub-permille agreement with a frequency-based reference, down to ring-down times of 4~$\mu$s. CRDS thereby combines benchmark accuracy with the simplicity that has made it widespread. For molecular spectroscopy, this means that tests of \textit{ab initio} theory, and the reference data on which atmospheric science and gas metrology rely, need no longer depend on a handful of specialized laboratories. Any laboratory, research or industrial, already running conventional CRDS has, in effect, already built an instrument capable of this accuracy. What it needs is only a characterization of a response that has been present in the instrument all along.

\section*{Materials and Methods}

\subsection*{Transfer-function measurement and model}
\label{appTF}

In the transfer-function measurement described in the main text (Eq.~\ref{eq1}), potential static nonlinearities in the intensity and beat-note amplitude measurements must be considered. Nonlinearities affecting the DC intensity measurements can be eliminated by keeping \(I_1^m\) and \(I_2^m\) constant while varying \(f_B\), since any resulting bias is then frequency independent and affects only the overall scale of \(|H(f_B)|\), which is removed by normalization. In contrast, static nonlinearities affecting the beat-note amplitude measurement become embedded in the measured transfer function, modifying its shape depending on the beat-note amplitude. Consequently, the experimentally determined \(|H(f_B)|\) represents the frequency response of the detection chain together with the residual effects of static nonlinearities present under the specific measurement conditions. To account for such level-dependent effects over the range of signal levels covered by the ring-down transients, the transfer-function parameters are refined within the multi-ring-down fitting procedure (see the ``Multi-ring-down refinement'' section).

The measured transfer function is modeled as the weighted sum of two low-pass components: a first-order component, which describes the slow part of the response and has a cutoff frequency far below the detection bandwidth, and a phenomenological higher-order resonant component, which describes the fast part, including a resonance near the upper edge of the detection band and the steep decrease above it, which limits the detection bandwidth. This fast response arises from detector electronics, parasitic circuit elements, and possible impedance mismatches. The resulting transfer-function model, normalized to \(H(0)=1\), is

\begin{equation}
\label{eq2}
H(f)=\frac{1}{1+A_r}\left[\frac{1}{1+i\frac{f}{f_c}}+\frac{A_r}{1-\left(\frac{f}{f_r}\right)^{n_r}+i\frac{f}{Q_r f_r}}\right],
\end{equation}
where \(f_c\) is the cutoff frequency of the first-order component, and \(f_r\), \(Q_r\), and \(n_r\) denote the resonance frequency, quality factor, and effective order of the resonant component, respectively (all positive). The parameter \(A_r\) is the ratio of the weights of the resonant and first-order components, which therefore carry fractions \(A_r/(1+A_r)\) and \(1/(1+A_r)\) of the signal, respectively (see the ``Fast and slow components of the detection response'' section). The functional form of the resonant term is based on the standard frequency response of a second-order system \cite{Franco2015}, with \(n_r\) introduced here as an effective parameter to account phenomenologically for the observed higher-order response. For configurations \(D_1\)--\(D_4\), \(f_c\approx 82\), \(83\), \(43\), and \(37\)~kHz and \(f_r\approx 5.35\), \(5.01\), \(5.63\), and \(5.75\)~MHz, respectively, so that the response decreases gradually above approximately \(10\)~kHz, shows a resonance near \(5\)~MHz, and drops steeply above approximately \(6\)~MHz. The detection bandwidth is therefore not a single model parameter but is set mainly by \(f_r\), \(Q_r\), and \(n_r\), whereas the ring-down signals measured here, whose spectra extend up to a few hundred kilohertz (shaded areas in Fig.~\ref{Fig_TF_plot_v2_selected2}A), are affected mainly by the first-order component with cutoff frequency \(f_c\). As shown by the black curve in Fig.~\ref{Fig_TF_plot_v2_selected2}A, the \(|H(f)|\) model accurately reproduces the measured transfer functions. The beat-note amplitude yields only the magnitude \(|H(f)|\). The phase response required by the time-domain forward model is therefore taken from the complex model of Eq.~\ref{eq2}, fitted to the measured magnitude and refined by the multi-ring-down fit (see the ``Multi-ring-down refinement'' section). The residuals of this fit are consistent with statistical noise, indicating that the model also reproduces the phase adequately over the frequency range spanned by the ring-down signals. An independent test is provided by the agreement of the TF-CRDS line areas with the CMDS reference for all truncation times.

\subsection*{Forward-model implementation}
\label{appFM}

This section details the numerical implementation of the forward model used in TF-CRDS to fit ring-down transients,
\begin{equation}
\label{eq3}
S_{\mathrm{fit}}(t)
=
\mathcal{F}^{-1}
\left\{
\mathcal{F}\left[S_{\mathrm{ed}}(t)\right]H(f)
\right\}.
\end{equation}
Here, we use the convention \(\mathcal{F}[g](f)=\int_{-\infty}^{\infty}g(t)\,e^{-i2\pi ft}\,dt\) and \(\mathcal{F}^{-1}[G](t)=\int_{-\infty}^{\infty}G(f)\,e^{i2\pi ft}\,df\) for the Fourier transform and its inverse. This convention fixes the sign of the imaginary terms in Eq.~\ref{eq2}, which is evaluated for \(f\geq0\) and extended to negative frequencies by \(H(-f)=H^{*}(f)\), so that the impulse response is real for any \(n_r>0\). If the exponential decay function $S_{\mathrm{ed}}(t)$ were defined only within the experimental time window $0\leq t\leq T$, evaluating Eq.~\ref{eq3} would introduce distortions at the edges of the modeled transient due to the implicit periodicity of the discrete Fourier transform. To minimize these well-known artifacts, $S_{\mathrm{ed}}(t)$ is extended by an additional interval of length $T$ on both sides, resulting in the interval $-T\leq t\leq2T$. Within this interval, $S_{\mathrm{ed}}(t)$ is defined as
\begin{equation}
\label{eq4}
S_{\mathrm{ed}}(t)=
\begin{cases}
b+B, & -T\leq t<0,\\
b+B\,e^{-t/\tau}, & 0\leq t\leq2T,
\end{cases}
\end{equation}
where $\tau$ is the ring-down time constant, while $b$ and $B$ are the offset and amplitude, respectively. The initial portion of the experimental ring-down up to a cutoff time $t_{\mathrm{cut}}$ is excluded from the fitting procedure to remove early-time ring-down distortions, such as those caused by delays between data acquisition and laser extinction, that are not accounted for by the forward model. The resulting $S_{\mathrm{fit}}(t)$ model is therefore compared with the experimental transient only over $t_{\mathrm{cut}} < t \leq T$.

Alternatively, the detection response could be compensated by deconvolution, i.e., dividing the Fourier spectrum of the measured signal by $H(f)$. However, this may amplify frequency-dependent noise and therefore requires noise-suppression techniques (e.g., Wiener filtering), which in turn may introduce systematic distortions. Deconvolution also prevents a consistent refinement of the transfer function, because the corrected data, including their amplified and correlated noise, change with the transfer-function parameters. Refining the transfer-function parameters by minimizing the residuals of the subsequent exponential fit would then compare fits to different data sets and would be biased toward parameters that amplify the noise least. In the forward model, by contrast, the measured data remain fixed and only the model changes, so that the transfer-function parameters can be refined together with the ring-down parameters (see the ``Multi-ring-down refinement'' section).

\subsection*{Fast and slow components of the detection response}
\label{appFS}

The transfer function of Eq.~\ref{eq2} can be written as the weighted sum of two parallel components of the detection response,
\begin{equation}
\label{eq5}
H(f)=\xi_{\mathrm{S}}\,H_{\mathrm{S}}(f)+\xi_{\mathrm{F}}\,H_{\mathrm{F}}(f),
\end{equation}
with the first-order and resonant low-pass terms
\begin{subequations}
\label{eq6}
\begin{gather}
H_{\mathrm{S}}(f)=\frac{1}{1+i\frac{f}{f_c}},\\
H_{\mathrm{F}}(f)=\frac{1}{1-\left(\frac{f}{f_r}\right)^{n_r}+i\frac{f}{Q_r f_r}},
\end{gather}
\end{subequations}
and the weights \(\xi_{\mathrm{S}}=1/(1+A_r)\) and \(\xi_{\mathrm{F}}=A_r/(1+A_r)=1-\xi_{\mathrm{S}}\). Since \((f/f_r)^{n_r}\) and \(f/(Q_r f_r)\) vanish at \(f=0\) for any \(n_r>0\) and \(Q_r\), \(H_{\mathrm{S}}(0)=H_{\mathrm{F}}(0)=1\), so that both components transmit the signal at low frequencies and \(H(0)=\xi_{\mathrm{S}}+\xi_{\mathrm{F}}=1\). The first-order term describes the slow component of the response, which relaxes with the time constant \(\tau_c=1/(2\pi f_c)\); its impulse response is \(h_{\mathrm{S}}(t)=\tau_c^{-1}e^{-t/\tau_c}\) for \(t\geq0\). The resonant term describes the fast component, which follows the signal within a fraction of a microsecond and sets the detection bandwidth.

\begin{figure}[t]
\centering
\includegraphics[width=0.47\textwidth]{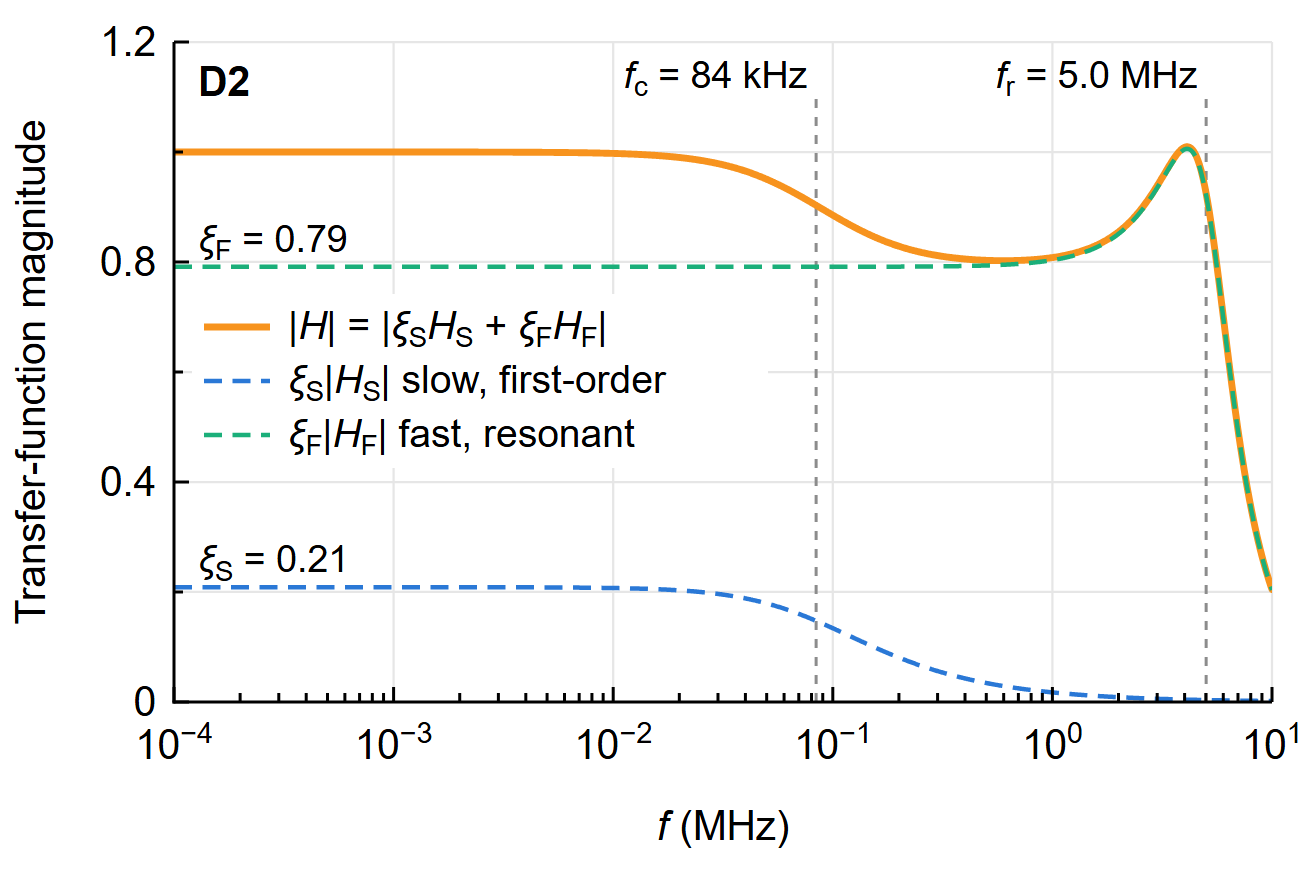}
\caption{\textbf{Fast and slow components of the detection transfer function.} Magnitude of the refined transfer function of the $D_2$ configuration (orange, as in Fig.~\ref{Fig_TF_plot_v2_selected2}A) and of its slow, first-order component $\xi_{\mathrm{S}}|H_{\mathrm{S}}|$ (dashed blue) and fast, resonant component $\xi_{\mathrm{F}}|H_{\mathrm{F}}|$ (dashed green), Eqs.~\ref{eq5} and \ref{eq6}. Well below $f_c$, the two components add up to unity. Above $f_c$, the slow component no longer follows the signal, and $|H|$ drops toward the level $\xi_{\mathrm{F}}$ of the fast component, which it then follows through the resonance near $f_r$ and the steep decrease above it. Because $|H|$ is the magnitude of the complex sum of the two components, it lies slightly below the sum of their magnitudes near $f_c$, where the response of the slow component is delayed in phase relative to that of the fast one (by $45^{\circ}$ at $f_c$).} \label{Fig_TF_components}
\end{figure}

For \(f\gg f_c\), \(H_{\mathrm{S}}(f)\approx-if_c/f\) and \(|H_{\mathrm{S}}(f)|\approx f_c/f\ll1\), so that the slow component no longer follows the signal, whereas for \(f\ll f_r\), \(H_{\mathrm{F}}(f)\approx1\). In the range \(f_c\ll f\ll f_r\), only the fast component therefore contributes, and \(|H(f)|\approx\xi_{\mathrm{F}}=1-\xi_{\mathrm{S}}\): the response forms a plateau between the low-frequency drop near \(f_c\) and the resonance near \(f_r\), lying below unity by the weight of the slow component (Fig.~\ref{Fig_TF_components}). The plateaus observed in the measured transfer functions (Fig.~\ref{Fig_TF_plot_v2_selected2}A) lie at the levels expected from the fitted weights, \(\xi_{\mathrm{S}}=15.4\%\), 20.9\%, 27.8\%, and 26.1\% for \(D_1\)--\(D_4\), respectively.

We now examine how the two components affect the recorded ring-down. The decay of Eq.~\ref{eq4} can be written as \(S_{\mathrm{ed}}(t)=b+B\,x(t)\), with \(x(t)=1\) for \(t<0\) and \(x(t)=e^{-t/\tau}\) for \(t\geq0\). The responses \(y_{\mathrm{S}}(t)\) and \(y_{\mathrm{F}}(t)\) of the two components to \(x(t)\) obey the time-domain equations corresponding to Eq.~\ref{eq6},
\begin{subequations}
\label{eq7}
\begin{gather}
\tau_c\,y'_{\mathrm{S}}+y_{\mathrm{S}}=x(t),\label{eq7a}\\
\frac{1}{\omega_r^{2}}\,y''_{\mathrm{F}}+\frac{1}{Q_r\omega_r}\,y'_{\mathrm{F}}+y_{\mathrm{F}}=x(t),\label{eq7b}
\end{gather}
\end{subequations}
where primes denote time derivatives and \(\omega_r=2\pi f_r\). Equation~\ref{eq7a} describes a first-order relaxation and Eq.~\ref{eq7b} a damped oscillator, written here for \(n_r=2\); a non-integer \(n_r\) modifies the response only near and above \(f_r\). For \(t\geq0\), the solution of each equation is the sum of a forced response, which has the shape of the input \(e^{-t/\tau}\) and differs from it only in amplitude, and a free response, which solves the equation with \(x=0\) and decays on the component's own time scale. The amplitudes of the free responses follow from the steady state before the light is switched off, in which \(y_{\mathrm{S}}=y_{\mathrm{F}}=1\) and neither changes in time. For the slow component this gives
\begin{equation}
\label{eqyL}
y_{\mathrm{S}}(t)=\frac{\tau}{\tau-\tau_c}\,e^{-t/\tau}-\frac{\tau_c}{\tau-\tau_c}\,e^{-t/\tau_c}.
\end{equation}
For the fast component, the forced response equals \(e^{-t/\tau}\) to within a relative correction of about \(1/(2\pi Q_r f_r\tau)\), i.e., about 1\% for \(f_r\approx5\)~MHz and the shortest ring-down time of \(4~\mu\mathrm{s}\), and the free response, a damped oscillation of similarly small amplitude, decays on the time scale \(Q_r/(\pi f_r)\), of the order of 100~ns. Neglecting these two small effects, the recorded transient \(S_{\mathrm{fit}}=b+B\,(\xi_{\mathrm{S}}y_{\mathrm{S}}+\xi_{\mathrm{F}}y_{\mathrm{F}})\) reduces to a sum of two exponential terms with time constants \(\tau\) and \(\tau_c\),
\begin{equation}
\label{eq8}
S_{\mathrm{fit}}(t)\approx b+B\left[\left(\xi_{\mathrm{F}}+\xi_{\mathrm{S}}\frac{\tau}{\tau-\tau_c}\right)e^{-t/\tau}-\xi_{\mathrm{S}}\frac{\tau_c}{\tau-\tau_c}\,e^{-t/\tau_c}\right].
\end{equation}

\begin{figure*}[t]
\centering
\includegraphics[width=1\textwidth]{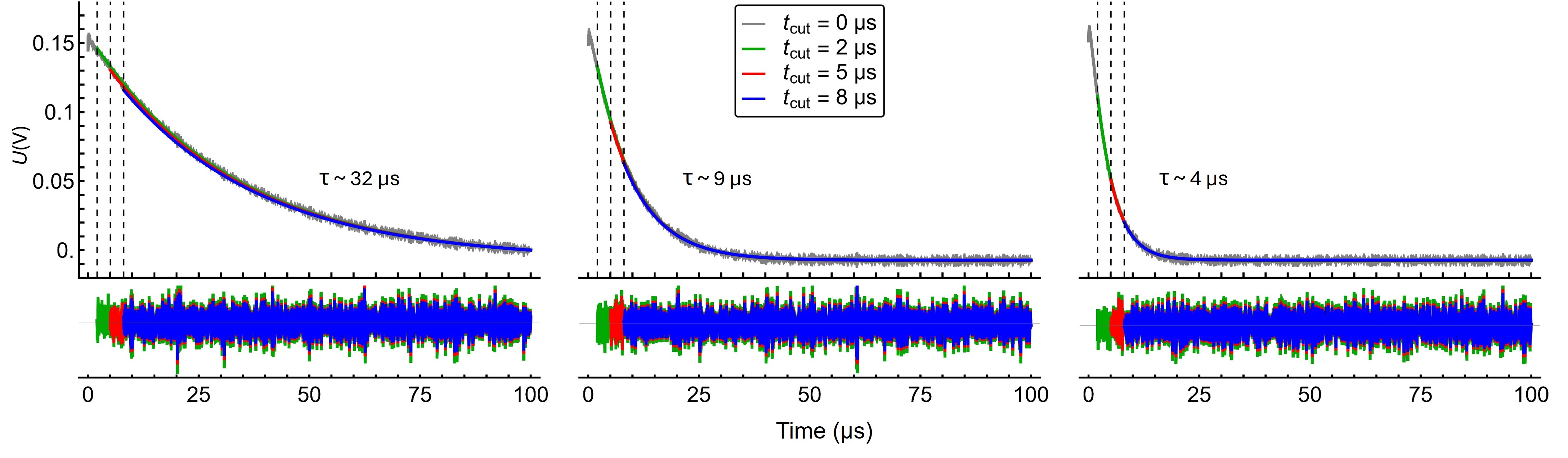}
\caption{\textbf{Multi-ring-down fit used to refine the effective transfer function.} Example of three ring-down transients used in the multi-ring-down fit to determine the effective $H(f)$ for the $D_2$ detection configuration (orange curve in Fig.~\ref{Fig_TF_plot_v2_selected2}A). The complete fit comprised 36 truncated transients obtained from six ring-down measurements, each averaged over 100 acquisitions, with decay time constants ranging from $4$ to $32~\mu\mathrm{s}$ and six initial truncation times, $t_{\mathrm{cut}}=2$, 3, 4, 5, 6, and $8~\mu\mathrm{s}$. Only a subset of decay time constants and $t_{\mathrm{cut}}$ values is shown. Below, residuals of the $S_{\mathrm{fit}}(t)$ TF-CRDS model fit are shown; the RMS is $\sim0.15~\mathrm{mV}$.
} \label{Fig_Multi_ring_down_fit}
\end{figure*}

Equation~\ref{eq8} shows where the distortion comes from. The forced responses reproduce the cavity decay with an unchanged time constant \(\tau\) and change only its amplitude; the distortion of the recorded decay therefore arises entirely from the free responses. If the detection response consisted of the fast component alone, its free response would vanish within a fraction of a microsecond, and truncating the early part of the transient would remove the distortion completely; this is why truncation appears sufficient. The free response of the slow component, the second term in Eq.~\ref{eq8}, decays instead with \(\tau_c=1.9\)--\(4.3~\mu\mathrm{s}\) for \(D_1\)--\(D_4\), comparable to the truncation times (\(t_{\mathrm{cut}}=2\)--\(8~\mu\mathrm{s}\)) and to the shortest ring-down times. It therefore persists in the fitted part of the transient and makes a single-exponential fit overestimate \(\tau\), increasingly so for shorter ring-down times. For \(\tau=\tau_c\), Eq.~\ref{eq8} remains finite and reduces to
\begin{equation}
\label{eq9}
S_{\mathrm{fit}}(t)\approx b+B\left(1+\xi_{\mathrm{S}}\,\frac{t}{\tau}\right)e^{-t/\tau},
\end{equation}
so that the recorded transient decays more slowly than the cavity decay from the very beginning and approaches the cavity decay rate only gradually; no practical truncation removes the bias. This is close to the situation at the line center for \(D_4\), where \(\tau\approx4~\mu\mathrm{s}\) and \(\tau_c\approx4.3~\mu\mathrm{s}\).

This analysis is illustrative. Bi-exponential models of this kind have been used before to account for a finite detector response time, with the whole response described by a single time constant~\cite{Gong2007}, which corresponds to \(\xi_{\mathrm{S}}=1\) and \(\xi_{\mathrm{F}}=0\) in Eq.~\ref{eq8}, i.e., to a response without a fast component. Such analytical expressions are, however, approximations. Equation~\ref{eq8} takes the forced response of the fast component as \(e^{-t/\tau}\), neglecting the correction of about 1\% to its amplitude for the shortest ring-down times. This correction changes the relative weights of the two terms in Eq.~\ref{eq8}; for the strongest distortions observed here, where the conventional and TF-CRDS time constants differ by up to 16\%, its effect on the retrieved \(\tau\) already exceeds the sub-permille level. A detection chain may, moreover, contain more than one slow component, each adding its own free response; an expression derived for a single slow component, such as Eq.~\ref{eq8}, would then no longer describe the recorded transient. The most reliable approach is therefore to measure the transfer function of each detection configuration and to use it in the forward model of Eq.~\ref{eq3}, which applies to any transfer function; all TF-CRDS results reported here were obtained in this way.

\subsection*{Multi-ring-down refinement}
\label{appMRDF}

This section describes the joint-fitting procedure used to refine the effective transfer function $H(f)$ referenced in the main text. To determine the optimal parameters of the effective transfer function $H(f)$, an $S_{\mathrm{fit}}(t)$-based multi-ring-down fit combines transients with different decay time constants, as they provide sensitivity to different spectral regions of the detection transfer function. Moreover, for each time constant, multiple versions of the same ring-down transient with different initial truncation times $t_{\mathrm{cut}}$ are included in the fit, making the effective $H(f)$ model also sensitive to potential static detection nonlinearities. In the absence of static nonlinearities, the same $H(f)$ would describe transients with all truncation times. Static nonlinearities can nevertheless produce different distortions for different $t_{\mathrm{cut}}$ values, since each truncation time corresponds to a different signal amplitude. Requiring a single $H(f)$ to fit all such transients effectively incorporates these effects into the transfer function, yielding an effective model that captures both the frequency-dependent response and residual level-dependent static nonlinearities within the range of signal amplitudes spanned by the fitted transients. The multi-ring-down fit also accounts for changes in the effective $H(f)$ arising from variations in the experimental conditions, such as changes in the beam alignment on the detector.

Figure~\ref{Fig_Multi_ring_down_fit} shows a subset of ring-down transients used in the multi-ring-down fit to determine the effective $H(f)$ for the $D_2$ detection configuration (orange curve in Fig.~\ref{Fig_TF_plot_v2_selected2}A). The fit included 36 truncated transients. Only the cutoff frequency $f_c$ of $H(f)$ was optimized simultaneously for all signals; the remaining parameters were fixed to values obtained directly from fitting the measured transfer function, to avoid numerical correlations among the transfer-function parameters in the joint fit. For each group of transients with the same decay time constant, the ring-down amplitude, decay time, and baseline were fitted jointly. The residuals from the global fit of all ring-down transients using the $S_{\mathrm{fit}}(t)$ model were consistent with statistical noise, with no systematic deviations observed. The $f_c$ obtained from this fit differs slightly from that obtained by directly fitting $H(f)$ to the experimental transfer function, likely due to the beam-alignment and residual static-nonlinearity effects discussed above. However, applying the same procedure to the $D_4$ configuration yielded excellent agreement between the effective $H(f)$ and the experimental transfer function within statistical uncertainties; compare the black and orange curves in Fig.~\ref{Fig_TF_plot_v2_selected2}A.

\subsection*{CRDS and CMDS spectroscopy setup}
\label{appSE}

\begin{figure}[t]
\centering
\includegraphics[width=0.47\textwidth]{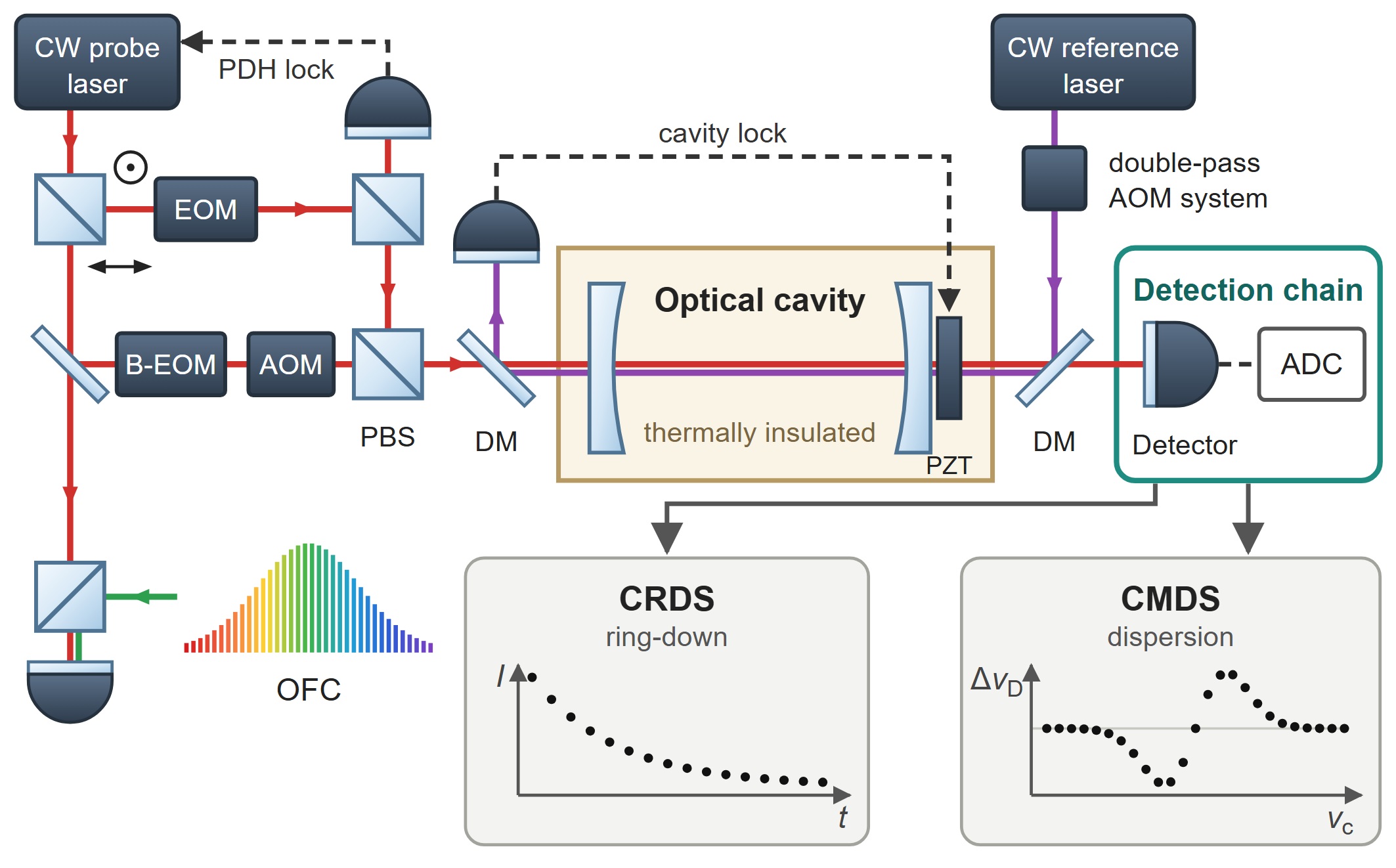}
\caption{\textbf{Experimental setup for simultaneous CRDS and CMDS measurements.} The CW probe laser is frequency-locked to a mode of the high-finesse optical cavity by the Pound--Drever--Hall (PDH) technique, using one of two orthogonally polarized branches of its beam. In the other branch, the tunable first-order sideband generated by a broadband electro-optic modulator (B-EOM) is tuned to consecutive cavity modes across the absorption line: the mode positions yield the CMDS spectrum, and ring-down decays, initiated by an acousto-optic modulator (AOM), yield the CRDS spectrum. The cavity length is stabilized with a piezoelectric actuator (PZT) to the iodine-stabilized Nd:YAG reference laser, whose light passes a double-pass AOM system and enters the cavity through dichroic mirrors (DM). The absolute laser frequency is measured with an optical frequency comb (OFC). The detection chain, investigated in four configurations, records the signals for both techniques. CW, continuous wave; EOM, electro-optic modulator; PBS, polarizing beam splitter; ADC, analog-to-digital converter.} \label{Fig_Spectroscopy_experiment}
\end{figure}

The experimental setup, shown in Fig.~\ref{Fig_Spectroscopy_experiment}, enables measurements using both CMDS and CRDS methods. A tunable external cavity diode laser (1520--1630~nm) is frequency locked to a high-finesse optical cavity containing the molecular sample using the Pound--Drever--Hall technique \cite{Drever1983}. The orthogonally polarized component of the laser beam serves as the probe beam and is sent through a broadband electro-optic modulator (B-EOM), which generates optical sidebands. The first-order sideband is used to scan consecutive cavity modes by tuning the modulation frequency over a 20~GHz range.

Precise measurements of the cavity mode profiles enable accurate determination of both the resonance linewidths and center frequencies \cite{Cygan2013,Cygan2015,Cygan2019}. Molecular dispersion shifts the cavity mode frequencies, and the CMDS signal is obtained by measuring these frequency shifts relative to the laser-locking mode, located outside the absorption feature. Owing to its purely frequency-domain nature, CMDS provides high accuracy, robustness against detector bandwidth limitations and static nonlinearities of the detection system \cite{Cygan2021,Cygan2025}, and enables direct SI frequency traceability, making it a suitable reference method for validation of CRDS measurements.

For CRDS measurements, after determining the cavity mode position, the laser light is switched off using an acousto-optic modulator (AOM), and the subsequent decay of the intracavity intensity is recorded. The cavity has a free spectral range of approximately 203~MHz and a finesse of $F=41\,000$. Its optical path length is actively stabilized to an iodine-stabilized Nd:YAG laser (1064~nm), resulting in an absolute frequency stability of the cavity modes below 10~kHz and a relative laser-to-cavity linewidth below 9~Hz \cite{Cygan2016,Cygan2025}. The absolute frequency of the cavity-locked laser beam is determined using an optical frequency comb (OFC). The cavity is thermally insulated, providing temperature stability and uniformity along the gas cell within 30~mK \cite{Bielska2022,Lisak2025}. The gas pressure is measured using capacitance manometers with a relative uncertainty of 0.05\% in the investigated pressure range. All frequency sources controlling the AOMs, EOMs, and the OFC reference are stabilized to a hydrogen maser frequency standard \cite{Cygan2016}. The laser-to-cavity lock, the cavity length stabilization, and the optical frequency comb serve the simultaneous CMDS reference and the frequency axis of the spectra; the ring-down decay itself begins after the probe light is switched off, and neither its analysis nor the transfer-function model relies on these elements. Further details on the experimental setup and its implementation can be found in Refs.~\cite{Cygan2015,Cygan2019,Cygan2025}.

In this work, the setup was used to measure the \(^{12}\mathrm{C}^{16}\mathrm{O}\) \(R(23)\) transition of the \((3-0)\) vibrational band at a pressure of 5~Torr (0.67~kPa) and a temperature of 296~K using both CMDS and CRDS techniques. The CO line intensity \(\left(8.0603(70)\times10^{-25}~\mathrm{cm}~\mathrm{molecule}^{-1}\right)\) \cite{Cygan2025} has previously been determined at the sub-permille level by several independent studies \cite{Bielska2022,CCQMP229,Cygan2025}. The investigated absorption range covered ring-down time constants from 32~\(\mu\mathrm{s}\) down to 4~\(\mu\mathrm{s}\). Under these conditions, the absorption remains linear. Saturation in the (3--0) band of CO has been observed for lines more than 20 times stronger than \(R(23)\), at pressures of 0.3--4~Pa rather than the 0.67~kPa used here, and in a cavity with a finesse of 390\,000, an order of magnitude higher than in the present setup~\cite{Bielska2021}. Each of these differences places the present measurements further from saturation.

Four detection configurations were investigated, combining an analog-to-digital converter (National Instruments PCI-5122, 100~MHz nominal bandwidth) with two InGaAs photodetectors of the same model (Newport 2053-FS, 10~MHz nominal bandwidth), operated at two different transimpedance gain settings (100 and 300). The detectors were nominally identical but differed in age and usage history; the older and newer units are referred to as detector~1 and detector~2, respectively. The configurations were labeled $D_1$ (detector~1, gain~100), $D_2$ (detector~1, gain~300), $D_3$ (detector~2, gain~100), and $D_4$ (detector~2, gain~300). Although detector~2 exhibited lower noise levels, the corresponding detection configurations showed larger transfer-function distortions than those based on detector~1, demonstrating that noise performance alone is not sufficient to assess detector suitability for high-accuracy cavity spectroscopy measurements.

In CRDS, ring-down transients were analyzed using a single-exponential decay model, whereas in TF-CRDS they were fitted with the \(S_{\mathrm{fit}}(t)\) forward model incorporating the effective transfer function \(H(f)\) obtained from the multi-ring-down refinement. The line-shape analysis employs a speed-dependent hard-collision profile (SDHCP) with a quadratic approximation of speed dependence \cite{Lance1997,Ngo2013,Wcislo2025}, recommended for high-accuracy atmospheric spectroscopy \cite{Tennyson2014}, thereby minimizing contributions from line-shape modeling errors.

\bibliography{bibfile}

@PREAMBLE{
 "\providecommand{\noopsort}[1]{}" 
 # "\providecommand{\singleletter}[1]{#1}%" 
}

@Article{ODell2018,
AUTHOR = {O'Dell, C. W. and Eldering, A. and Wennberg, P. O. and Crisp, D. and Gunson, M. R. and Fisher, B. and Frankenberg, C. and Kiel, M. and Lindqvist, H. and Mandrake, L. and Merrelli, A. and Natraj, V. and Nelson, R. R. and Osterman, G. B. and Payne, V. H. and Taylor, T. E. and Wunch, D. and Drouin, B. J. and Oyafuso, F. and Chang, A. and McDuffie, J. and Smyth, M. and Baker, D. F. and Basu, S. and Chevallier, F. and Crowell, S. M. R. and Feng, L. and Palmer, P. I. and Dubey, M. and Garc\'{\i}a, O. E. and Griffith, D. W. T. and Hase, F. and Iraci, L. T. and Kivi, R. and Morino, I. and Notholt, J. and Ohyama, H. and Petri, C. and Roehl, C. M. and Sha, M. K. and Strong, K. and Sussmann, R. and Te, Y. and Uchino, O. and Velazco, V. A.},
TITLE = {Improved retrievals of carbon dioxide from Orbiting Carbon Observatory-2 with the version 8 ACOS algorithm},
JOURNAL = {Atmospheric Measurement Techniques},
VOLUME = {11},
YEAR = {2018},
NUMBER = {12},
PAGES = {6539--6576},
URL = {https://amt.copernicus.org/articles/11/6539/2018/},
DOI = {10.5194/amt-11-6539-2018}
}

@article{Basu2011,
author = {Basu, Sourish and Houweling, Sander and Peters, Wouter and Sweeney, Colm and Machida, Toshinobu and Maksyutov, Shamil and Patra, Prabir K. and Saito, Ryu and Chevallier, Frederic and Niwa, Yosuke and Matsueda, Hidekazu and Sawa, Yousuke},
title = {The seasonal cycle amplitude of total column CO2: Factors behind the model-observation mismatch},
journal = {Journal of Geophysical Research: Atmospheres},
volume = {116},
number = {D23},
pages = {1-14},
doi = {https://doi.org/10.1029/2011JD016124},
url = {https://agupubs.onlinelibrary.wiley.com/doi/abs/10.1029/2011JD016124},
year = {2011}
}

@article{Wunch2011,
    author = {Wunch, Debra and Toon, Geoffrey C. and Blavier, Jean-François L. and Washenfelder, Rebecca A. and Notholt, Justus and Connor, Brian J. and Griffith, David W. T. and Sherlock, Vanessa and Wennberg, Paul O.},
    title = {The Total Carbon Column Observing Network},
    journal = {Philosophical Transactions of the Royal Society A: Mathematical, Physical and Engineering Sciences},
    volume = {369},
    number = {1943},
    pages = {2087-2112},
    year = {2011},
    month = {05},
    doi = {10.1098/rsta.2010.0240},
    url = {https://doi.org/10.1098/rsta.2010.0240}
}

@article{Bagdonaite2015,
  title = {Constraint on a Varying Proton-Electron Mass Ratio 1.5 Billion Years after the Big Bang},
  author = {Bagdonaite, J. and Ubachs, W. and Murphy, M. T. and Whitmore, J. B.},
  journal = {Phys. Rev. Lett.},
  volume = {114},
  issue = {7},
  pages = {071301},
  numpages = {6},
  year = {2015},
  month = {Feb},
  publisher = {American Physical Society},
  doi = {10.1103/PhysRevLett.114.071301},
  url = {https://link.aps.org/doi/10.1103/PhysRevLett.114.071301}
}

@article{Ubachs2018,
  title = {Constraint on a Varying Proton-Electron Mass Ratio 1.5 Billion Years after the Big Bang},
  author = {Ubachs, W.},
  journal = {Space Sci. Rev.},
  volume = {3},
  pages = {1},
  year = {2018},
  doi = {10.1007/s11214-017-0432-y}
}

@article{Webb2011,
  title = {Indications of a Spatial Variation of the Fine Structure Constant},
  author = {Webb, J. K. and King, J. A. and Murphy, M. T. and Flambaum, V. V. and Carswell, R. F. and Bainbridge, M. B.},
  journal = {Phys. Rev. Lett.},
  volume = {107},
  issue = {19},
  pages = {191101},
  numpages = {5},
  year = {2011},
  month = {Oct},
  publisher = {American Physical Society},
  doi = {10.1103/PhysRevLett.107.191101},
  url = {https://link.aps.org/doi/10.1103/PhysRevLett.107.191101}
}

@article{HITRAN2024,
title = {The HITRAN2024 molecular spectroscopic database},
journal = {Journal of Quantitative Spectroscopy and Radiative Transfer},
volume = {353},
pages = {109807},
year = {2026},
issn = {0022-4073},
doi = {https://doi.org/10.1016/j.jqsrt.2026.109807},
url = {https://www.sciencedirect.com/science/article/pii/S0022407326000014},
author = {I.E. Gordon and L.S. Rothman and R.J. Hargreaves and F.M. Gomez and T. Bertin and C. Hill and R.V. Kochanov and Y. Tan and P. Wcisło and V. Yu. Makhnev and P.F. Bernath and M. Birk and V. Boudon and A. Campargue and A. Coustenis and B.J. Drouin and R.R. Gamache and J.T. Hodges and D. Jacquemart and E.J. Mlawer and A.V. Nikitin and V.I. Perevalov and M. Rotger and S. Robert and J. Tennyson and G.C. Toon and H. Tran and V.G. Tyuterev and E.M. Adkins and A. Barbe and D.M. Bailey and K. Bielska and L. Bizzocchi and T.A. Blake and C.A. Bowesman and P. Cacciani and P. Čermák and A.G. Császár and L. Denis and S.C. Egbert and O. Egorov and A. Yu. Ermilov and A.J. Fleisher and H. Fleurbaey and A. Foltynowicz and T. Furtenbacher and M. Germann and E.R. Guest and J.J. Harrison and J.-M. Hartmann and A. Hjältén and S.-M. Hu and X. Huang and T.J. Johnson and H. Jóźwiak and S. Kassi and M.V. Khan and F. Kwabia-Tchana and T.J. Lee and D. Lisak and A.-W. Liu and O.M. Lyulin and N.A. Malarich and L. Manceron and A.A. Marinina and S.T. Massie and J. Mascio and E.S. Medvedev and V.V. Meshkov and G. Ch. Mellau and M. Melosso and S.N. Mikhailenko and D. Mondelain and H.S.P. Müller and M. O’Donnell and A. Owens and A. Perrin and O.L. Polyansky and P.L. Raston and Z.D. Reed and M. Rey and C. Richard and G.B. Rieker and C. Röske and S.W. Sharpe and E. Starikova and N. Stolarczyk and A.V. Stolyarov and K. Sung and F. Tamassia and J. Terragni and V.G. Ushakov and S. Vasilchenko and B. Vispoel and K.L. Vodopyanov and G. Wagner and S. Wójtewicz and S.N. Yurchenko and N.F. Zobov}
}

@article{Tennyson2012,
    author = {Tennyson, Jonathan and Yurchenko, Sergei N.},
    title = {ExoMol: molecular line lists for exoplanet and other atmospheres},
    journal = {Monthly Notices of the Royal Astronomical Society},
    volume = {425},
    number = {1},
    pages = {21-33},
    year = {2012},
    month = {09},
        issn = {0035-8711},
    doi = {10.1111/j.1365-2966.2012.21440.x},
    url = {https://doi.org/10.1111/j.1365-2966.2012.21440.x}
}

@ARTICLE{Canocchi2024,
       author = {{Canocchi}, G. and {Lind}, K. and {Lagae}, C. and {Pietrow}, A.~G.~M. and {Amarsi}, A.~M. and {Kiselman}, D. and {Andriienko}, O. and {Hoeijmakers}, H.~J.},
        title = "{3D non-LTE modeling of the stellar center-to-limb variation for transmission spectroscopy studies. Na I D and K I resonance lines in the Sun}",
    journal = {Astronomy \& Astrophysics},
         year = 2024,
        month = mar,
       volume = {683},
          eid = {A242},
        pages = {A242},
          doi = {10.1051/0004-6361/202347858},
       adsurl = {https://ui.adsabs.harvard.edu/abs/2024A&A...683A.242C}
}

@article{Polyansky2018,
    author = {Polyansky, Oleg L and Kyuberis, Aleksandra A and Zobov, Nikolai F and Tennyson, Jonathan and Yurchenko, Sergei N and Lodi, Lorenzo},
    title = {ExoMol molecular line lists XXX: a complete high-accuracy line list for water},
    journal = {Monthly Notices of the Royal Astronomical Society},
    volume = {480},
    number = {2},
    pages = {2597-2608},
    year = {2018},
    month = {10},
    issn = {0035-8711},
    doi = {10.1093/mnras/sty1877},
    url = {https://doi.org/10.1093/mnras/sty1877}
}

@article{Sanchez2021,
   author = "Quintas-Sánchez, Ernesto and Dawes, Richard",
   title = "Spectroscopy and Scattering Studies Using Interpolated Ab Initio Potentials", 
   journal= "Annual Review of Physical Chemistry",
   year = "2021",
   volume = "72",
   number = "Volume 72, 2021",
   pages = "399-421",
   doi = "https://doi.org/10.1146/annurev-physchem-090519-051837"
  }

@article{Kowzan2020,
title = {Fully quantum calculations of the line-shape parameters for the Hartmann-Tran profile: A CO-Ar case study},
journal = {Journal of Quantitative Spectroscopy and Radiative Transfer},
volume = {243},
pages = {106803},
year = {2020},
issn = {0022-4073},
doi = {https://doi.org/10.1016/j.jqsrt.2019.106803},
author = {Grzegorz Kowzan and Piotr Wcisło and Michał Słowiński and Piotr Masłowski and Alexandra Viel and Franck Thibault}
}

@article{Zak2017,
title = {Room temperature line lists for CO2 symmetric isotopologues with ab initio computed intensities},
journal = {Journal of Quantitative Spectroscopy and Radiative Transfer},
volume = {189},
pages = {267-280},
year = {2017},
issn = {0022-4073},
doi = {https://doi.org/10.1016/j.jqsrt.2016.11.022},
author = {Emil J. Zak and Jonathan Tennyson and Oleg L. Polyansky and Lorenzo Lodi and Nikolay F. Zobov and Sergei A. Tashkun and Valery I. Perevalov}
}

@article{Hartmann2013,
  title = {$Ab\phantom{\rule{0.28em}{0ex}}\phantom{\rule{0.28em}{0ex}}initio$ calculations of the spectral shapes of CO${}_{2}$ isolated lines including non-Voigt effects and comparisons with experiments},
  author = {Hartmann, J.-M. and Tran, H. and Ngo, N. H. and Landsheere, X. and Chelin, P. and Lu, Y. and Liu, A.-W. and Hu, S.-M. and Gianfrani, L. and Casa, G. and Castrillo, A. and Lep\`ere, M. and Deli\`ere, Q. and Dhyne, M. and Fissiaux, L.},
  journal = {Phys. Rev. A},
  volume = {87},
  issue = {1},
  pages = {013403},
  numpages = {11},
  year = {2013},
  month = {Jan},
  publisher = {American Physical Society},
  doi = {10.1103/PhysRevA.87.013403}
}

@article{Birk2017,
title = {Accurate line intensities for water transitions in the infrared: Comparison of theory and experiment},
journal = {Journal of Quantitative Spectroscopy and Radiative Transfer},
volume = {203},
pages = {88-102},
year = {2017},
issn = {0022-4073},
doi = {https://doi.org/10.1016/j.jqsrt.2017.03.040},
url = {https://www.sciencedirect.com/science/article/pii/S0022407317300596},
author = {Manfred Birk and Georg Wagner and Joep Loos and Lorenzo Lodi and Oleg L. Polyansky and Aleksandra A. Kyuberis and Nikolai F. Zobov and Jonathan Tennyson}
}

@article{Rubin2022,
author = {Tom M. Rubin and Marian Sarrazin and Nikolai F. Zobov and Jonathan Tennyson and Oleg L. Polyansky},
title = {Sub-percent accuracy for the intensity of a near-infrared water line at 10,670 cm−1: experiment and analysis},
journal = {Molecular Physics},
volume = {120},
number = {19-20},
pages = {e2063769},
year = {2022},
publisher = {Taylor \& Francis},
doi = {10.1080/00268976.2022.2063769}
}

@article{Cygan2025,
    author = {Agata Cygan and Szymon W\'{o}jtewicz and Hubert J\'{o}\'{z}wiak and Grzegorz Kowzan and Nikodem Stolarczyk and Katarzyna Bielska and Piotr Wcis{\l}o and Roman Ciury{\l}o and Daniel Lisak}, 
    title = {Dispersive heterodyne cavity ring-down spectroscopy exploiting eigenmode frequencies for high-fidelity
measurements}, 
    journal        = {Sci. Adv.},
    year           = {2025},
    volume         = {11},
    number         = {5},
    pages          = {eadp8556},
    doi            = {10.1126/sciadv.adp8556},
}

@article{Li2025,
author = {Jin-Ke Li  and Jin Wang  and Rui-Heng Yin  and Qi Huang  and Yan Tan  and Chang-Le Hu  and Yu R. Sun  and Oleg L. Polyansky  and Nikolai F. Zobov  and Evgenii I. Lebedev  and Rainer Stosch  and Jonathan Tennyson  and Gang Li  and Shui-Ming Hu },
title = {Unprecedented accuracy in molecular line-intensity ratios from frequency-based measurements},
journal = {Science Advances},
volume = {11},
number = {38},
pages = {eadz6560},
year = {2025},
doi = {10.1126/sciadv.adz6560},
URL = {https://www.science.org/doi/abs/10.1126/sciadv.adz6560}
}

@article{Lisak2025,
  title = {Leveraging Resonant Frequencies of an Optical Cavity for Spectroscopic Measurement of Gas Temperature and Concentration},
  author = {Lisak, Daniel and D'Agostino, Vittorio and W\'ojtewicz, Szymon and Cygan, Agata and Gibas, Marcin and Wcis\l{}o, Piotr and Ciury\l{}o, Roman and Bielska, Katarzyna},
  journal = {Phys. Rev. Lett.},
  volume = {135},
  issue = {10},
  pages = {103201},
  numpages = {8},
  year = {2025},
  month = {Sep},
  publisher = {American Physical Society},
  doi = {10.1103/2jz1-dr5l},
  url = {https://link.aps.org/doi/10.1103/2jz1-dr5l}
}

@article{Bielska2022,
  title = {Subpromille Measurements and Calculations of CO (3--0) Overtone Line Intensities},
  author = {Bielska, Katarzyna and Kyuberis, Aleksandra A. and Reed, Zachary D. and Li, Gang and Cygan, Agata and Ciury\l{}o, Roman and Adkins, Erin M. and Lodi, Lorenzo and Zobov, Nikolay F. and Ebert, Volker and Lisak, Daniel and Hodges, Joseph T. and Tennyson, Jonathan and Polyansky, Oleg L.},
  journal = {Phys. Rev. Lett.},
  volume = {129},
  issue = {4},
  pages = {043002},
  numpages = {6},
  year = {2022},
  month = {Jul},
  publisher = {American Physical Society},
  doi = {10.1103/PhysRevLett.129.043002},
}

@article{Bielska2021,
author = {Katarzyna Bielska and Agata Cygan and Magdalena Konefa{\l} and Grzegorz Kowzan and Miko{\l}aj Zaborowski and Dominik Charczun and Szymon W\'{o}jtewicz and Piotr Wcis{\l}o and Piotr Mas{\l}owski and Roman Ciury{\l}o and Daniel Lisak},
journal = {Opt. Express},
number = {24},
pages = {39449--39460},
publisher = {Optica Publishing Group},
title = {Frequency-based dispersion Lamb-dip spectroscopy in a high finesse optical cavity},
volume = {29},
month = {Nov},
year = {2021},
url = {https://opg.optica.org/oe/abstract.cfm?URI=oe-29-24-39449},
doi = {10.1364/OE.443661}
}

@article{CCQMP229,
    author         = { J. T. Hodges and K. Bielska and M. Birk and R. Guo and G. Li and J. S. Lim and D. Lisak and Z. D. Reed and G. Wagner},
    title          = {International Comparison {CCQM-P229: Pilot} Study to Measure Absolute Line Intensities of Selected {$^{12}$C$^{16}$O} Transitions}, 
    journal        = {Metrologia},
    year           = {2025},
    volume         = {62},
    pages          = {08006},
    doi            = {10.1088/0026-1394/62/1A/08006},
    url            = {https://dx.doi.org/10.1088/0026-1394/62/1A/08006},
}

@article{Adkins2025,
title = {An accurate determination of O2 A-band line intensities through experiment and theory},
journal = {Journal of Quantitative Spectroscopy and Radiative Transfer},
volume = {338},
pages = {109412},
year = {2025},
issn = {0022-4073},
doi = {https://doi.org/10.1016/j.jqsrt.2025.109412},
url = {https://www.sciencedirect.com/science/article/pii/S0022407325000743},
author = {Erin M. Adkins and Sergei N. Yurchenko and Wilfrid Somogyi and Joseph T. Hodges}
}

@article{Fleisher2021,
  author  = {Fleisher, Adam J. and Yi, Hongming and Srivastava, Abneesh and Polyansky, Oleg L. and Zobov, Nikolai F. and Hodges, Joseph T.},
  title   = {Absolute $^{13}$C/$^{12}$C isotope amount ratio for Vienna Pee Dee Belemnite from infrared absorption spectroscopy},
  journal = {Nature Physics},
  volume  = {17},
  pages   = {889--893},
  year    = {2021},
  doi     = {10.1038/s41567-021-01226-y}
}

@article{Polyansky2015,
  title = {High-Accuracy ${\mathrm{CO}}_{2}$ Line Intensities Determined from Theory and Experiment},
  author = {Polyansky, Oleg L. and Bielska, Katarzyna and Ghysels, M\'elanie and Lodi, Lorenzo and Zobov, Nikolai F. and Hodges, Joseph T. and Tennyson, Jonathan},
  journal = {Phys. Rev. Lett.},
  volume = {114},
  issue = {24},
  pages = {243001},
  year = {2015},
  month = {Jun},
  doi = {10.1103/PhysRevLett.114.243001},
}

@article{OKeefe1988,
  author  = {O'Keefe, A. and Deacon, D. A. G.},
  title   = {Cavity ring-down optical spectrometer for absorption measurements using pulsed laser sources},
  journal = {Review of Scientific Instruments},
  volume  = {59},
  pages   = {2544--2551},
  year    = {1988},
  doi     = {10.1063/1.1139895}
}

@article{Romanini1997,
title = {CW cavity ring down spectroscopy},
journal = {Chem. Phys. Lett.},
volume = {264},
number = {3},
pages = {316-322},
year = {1997},
doi = {https://doi.org/10.1016/S0009-2614(96)01351-6},
author = {D. Romanini and A.A. Kachanov and N. Sadeghi and F. Stoeckel},
}

@article{Wojtewicz2011,
  title = {Line-shape study of self-broadened O${}_{2}$ transitions measured by Pound-Drever-Hall-locked frequency-stabilized cavity ring-down spectroscopy},
  author = {W\'ojtewicz, S. and Lisak, D. and Cygan, A. and Domys\l{}awska, J. and Trawi\ifmmode \acute{n}\else \'{n}\fi{}ski, R. S. and Ciury\l{}o, R.},
  journal = {Phys. Rev. A},
  volume = {84},
  issue = {3},
  pages = {032511},
  numpages = {9},
  year = {2011},
  month = {Sep},
  publisher = {American Physical Society},
  doi = {10.1103/PhysRevA.84.032511},
  url = {https://link.aps.org/doi/10.1103/PhysRevA.84.032511}
}

@article{Fleisher2019,
  title = {Twenty-Five-Fold Reduction in Measurement Uncertainty for a Molecular Line Intensity},
  author = {Fleisher, Adam J. and Adkins, Erin M. and Reed, Zachary D. and Yi, Hongming and Long, David A. and Fleurbaey, H\'el\`ene M. and Hodges, Joseph T.},
  journal = {Phys. Rev. Lett.},
  volume = {123},
  issue = {4},
  pages = {043001},
  numpages = {7},
  year = {2019},
  month = {Jul},
  publisher = {American Physical Society},
  doi = {10.1103/PhysRevLett.123.043001},
  url = {https://link.aps.org/doi/10.1103/PhysRevLett.123.043001}
}

@article{Cao2019,
  author  = {Cao, Zhensong and Li, Zhixin and Xu, Fei and Wu, Yongqian and Zhou, Zixin and Tong, Zhaomin and Ma, Weiguang and Zhu, Wenyue},
  title   = {Influence of Spatial Inhomogeneity of Detector Temporal Responses on the Spectral Fidelity in Continuous Wave Cavity Ringdown Spectroscopy},
  journal = {Sensors},
  volume  = {19},
  number  = {23},
  pages   = {5232},
  year    = {2019},
  doi     = {10.3390/s19235232}
}

@article{HITRAN2017,
title = {The HITRAN2016 molecular spectroscopic database},
journal = {Journal of Quantitative Spectroscopy and Radiative Transfer},
volume = {203},
pages = {3-69},
year = {2017},
issn = {0022-4073},
doi = {https://doi.org/10.1016/j.jqsrt.2017.06.038},
author = {I.E. Gordon and L.S. Rothman and C. Hill and R.V. Kochanov and Y. Tan and P.F. Bernath and M. Birk and V. Boudon and A. Campargue and K.V. Chance and B.J. Drouin and J.-M. Flaud and R.R. Gamache and J.T. Hodges and D. Jacquemart and V.I. Perevalov and A. Perrin and K.P. Shine and M.-A.H. Smith and J. Tennyson and G.C. Toon and H. Tran and V.G. Tyuterev and A. Barbe and A.G. Császár and V.M. Devi and T. Furtenbacher and J.J. Harrison and J.-M. Hartmann and A. Jolly and T.J. Johnson and T. Karman and I. Kleiner and A.A. Kyuberis and J. Loos and O.M. Lyulin and S.T. Massie and S.N. Mikhailenko and N. Moazzen-Ahmadi and H.S.P. Müller and O.V. Naumenko and A.V. Nikitin and O.L. Polyansky and M. Rey and M. Rotger and S.W. Sharpe and K. Sung and E. Starikova and S.A. Tashkun and J. Vander Auwera and G. Wagner and J. Wilzewski and P. Wcisło and S. Yu and E.J. Zak}
}

@article{Pachucki2025,
author = {Pachucki, Krzysztof and Komasa, Jacek},
title = {From First-Principles to Quantum Electrodynamics: Pushing the Limits of Theory with the Hydrogen Molecule},
journal = {Journal of Chemical Theory and Computation},
volume = {21},
number = {24},
pages = {12664-12673},
year = {2025},
doi = {10.1021/acs.jctc.5c01702}
}

@article{Stankiewicz2026,
  title = {Cavity-enhanced spectroscopy in the deep cryogenic regime for quantum sensing and metrology},
  author = {
    Stankiewicz, Kamil and Makowski, Marcin and Słowiński, Michał and Sołtys, Kamil L. and
    Bednarski, Bogdan and Jóźwiak, Hubert J. and Stolarczyk, Nikodem and Narożnik, Mateusz and
    Kierski, Dariusz and Wójtewicz, Szymon and Cygan, Agata and Kowzan, Grzegorz and
    Masłowski, Piotr and Piwiński, Mariusz and Lisak, Daniel and Wcisło, Piotr
  },
  journal = {Nature Physics},
  volume = {22},
  pages = {637--643},
  year = {2026},
  doi = {10.1038/s41567-026-03204-8},
  publisher = {Springer Nature}
}

@article{Hashiguchi2023,
title = {Accurate determination of line intensity of H2O near 7181 cm−1 : SI-traceable measurement using primary trace-moisture standard in N2 gas},
journal = {Journal of Quantitative Spectroscopy and Radiative Transfer},
volume = {311},
pages = {108784},
year = {2023},
issn = {0022-4073},
doi = {https://doi.org/10.1016/j.jqsrt.2023.108784},
url = {https://www.sciencedirect.com/science/article/pii/S0022407323003023},
author = {Koji Hashiguchi and Minami Amano and Agata Cygan and Daniel Lisak and Roman Ciuryło and Hisashi Abe}
}

@article{Castrillo2024,
author = {Antonio Castrillo and Muhammad Asad Khan and Eugenio Fasci and Vittorio D'Agostino and Stefania Gravina and Livio Gianfrani},
journal = {Optica},
number = {9},
pages = {1277--1284},
publisher = {Optica Publishing Group},
title = {Demonstration of record sensitivity for water vapor detection by means of comb-locked cavity ring-down spectroscopy},
volume = {11},
month = {Sep},
year = {2024},
url = {https://opg.optica.org/optica/abstract.cfm?URI=optica-11-9-1277},
doi = {10.1364/OPTICA.531464}
}

@article{Gotti2018,
  title = {Cavity-ring-down Doppler-broadening primary thermometry},
  author = {Gotti, Riccardo and Moretti, Luigi and Gatti, Davide and Castrillo, Antonio and Galzerano, Gianluca and Laporta, Paolo and Gianfrani, Livio and Marangoni, Marco},
  journal = {Phys. Rev. A},
  volume = {97},
  issue = {1},
  pages = {012512},
  numpages = {5},
  year = {2018},
  month = {Jan},
  publisher = {American Physical Society},
  doi = {10.1103/PhysRevA.97.012512},
  url = {https://link.aps.org/doi/10.1103/PhysRevA.97.012512}
}

@article{Cygan2015,
author = {Agata Cygan and Piotr Wcis{\l}o and Szymon W\'{o}jtewicz and Piotr Mas{\l}owski and Joseph T. Hodges and Roman Ciury{\l}o and Daniel Lisak},
journal = {Opt. Express},
number = {11},
pages = {14472--14486},
title = {One-dimensional frequency-based spectroscopy},
volume = {23},
year = {2015},
doi = {10.1364/OE.23.014472}
}

@article{Cygan2016,
    author = {Cygan, A. and W\'{o}jtewicz, S. and Kowzan, G. and Zaborowski, M. and Wcis{\l}o, P. and Nawrocki, J. and Krehlik, P. and \'{S}liwczyński, {\L}. and Lipi\'{n}ski, M. and Mas{\l}owski, P. and Ciury{\l}o, R. and Lisak, D.},
    title = "{Absolute molecular transition frequencies measured by three cavity-enhanced spectroscopy techniques}",
    journal = {J. Chem. Phys.},
    volume = {144},
    number = {21},
    pages = {214202},
    year = {2016},
    doi = {10.1063/1.4952651}
}

@article{Cygan2019,
author = {Agata Cygan and Piotr Wcis{\l}o and Szymon W\'{o}jtewicz and Grzegorz Kowzan and Miko{\l}aj Zaborowski and Dominik Charczun and Katarzyna Bielska and Ryszard S. Trawi\'{n}ski and Roman Ciury{\l}o and Piotr Mas{\l}owski and Daniel Lisak},
journal = {Opt. Express},
number = {15},
pages = {21810--21821},
title = {High-accuracy and wide dynamic range frequency-based dispersion spectroscopy in an optical cavity},
volume = {27},
year = {2019},
doi = {10.1364/OE.27.021810}
}

@book{Franco2015,
  author    = {Sergio Franco},
  title     = {Design with Operational Amplifiers and Analog Integrated Circuits},
  edition   = {4},
  publisher = {McGraw-Hill Education},
  address   = {New York},
  year      = {2015}
}

@article{Cygan2013,
author = {Agata Cygan and Daniel Lisak and Piotr Morzy\'{n}ski and Marcin Bober and Micha{\l} Zawada and Eugeniusz Pazderski and Roman Ciury{\l}o},
journal = {Opt. Express},
number = {24},
pages = {29744--29754},
title = {Cavity mode-width spectroscopy with widely tunable ultra narrow laser},
volume = {21},
year = {2013},
doi = {10.1364/OE.21.029744}
}

@article{Drever1983,
    author         = {R. W. P. Drever and J. L. Hall and F. V. Kowalski and J. Hough and G. M. Ford and A. J. Munley and H. Ward}, 
    title          = {Laser phase and frequency stabilization using an optical resonator}, 
    journal        = {Appl. Phys. B },
    year            = {1983},
    volume         = {31},
    pages          = {97-105},
    doi             = {10.1007/BF00702605}
}

@article{Huang2024,
author = {Q. Huang and Y. Tan and R.-H. Yin and Z.-L. Nie and J. Wang and S.-M. Hu},
title = {Line intensities of {CO} near 1560 nm measured with absorption and dispersion spectroscopy},
journal = {Metrologia},
volume = {61},
number = {},
pages = {065003},
year = {2024},
doi = {10.1088/1681-7575/ad7ec0}
}

@article{Cygan2021,
    author         = {Agata Cygan and Adam J. Fleisher and Roman Ciuryło and Keith A. Gillis and Joseph T. Hodges and Daniel Lisak}, 
    title          = {Cavity buildup dispersion spectroscopy}, 
    journal        = {Communic. Phys.},
    year            = {2021},
    volume         = {4},
    pages          = {14},
    doi             = {10.1038/s42005-021-00517-3}
}

@article{Lance1997,
title = {On the Speed-Dependent Hard Collision Lineshape Models: Application to C2H2Perturbed by Xe},
journal = {Journal of Molecular Spectroscopy},
volume = {185},
number = {2},
pages = {262-271},
year = {1997},
issn = {0022-2852},
doi = {https://doi.org/10.1006/jmsp.1997.7385},
url = {https://www.sciencedirect.com/science/article/pii/S0022285297973859},
author = {Benoit Lance and Ghislain Blanquet and Jacques Walrand and Jean-Pierre Bouanich}
}

@article{Ngo2013,
title = {An isolated line-shape model to go beyond the {Voigt} profile in spectroscopic databases and radiative transfer codes},
journal = {J. Quant. Spectrosc. Radiat. Transf.},
volume = {129},
pages = {89-100},
year = {2013},
doi = {10.1016/j.jqsrt.2013.05.034},
author = {N.H. Ngo and D. Lisak and H. Tran and J.-M. Hartmann}
}

@article{Tennyson2014,
title = {Recommended isolated-line profile for representing high-resolution spectroscopic transitions (IUPAC Technical Report)},
author = {Jonathan Tennyson and Peter F. Bernath and Alain Campargue and Attila G. Császár and Ludovic Daumont and Robert R. Gamache and Joseph T. Hodges and Daniel Lisak and Olga V. Naumenko and Laurence S. Rothman and Ha Tran and Nikolai F. Zobov and Jeanna Buldyreva and Chris D. Boone and Maria Domenica De Vizia and Livio Gianfrani and Jean-Michel Hartmann and Robert McPheat and Damien Weidmann and Jonathan Murray and Ngoc Hoa Ngo and Oleg L. Polyansky},
pages = {1931--1943},
volume = {86},
number = {12},
journal = {Pure Appl. Chem.},
doi = {doi:10.1515/pac-2014-0208},
year = {2014}
}

@article{Wcislo2025,
title = {New beyond-Voigt line-shape profile recommended for the HITRAN database},
journal = {Journal of Quantitative Spectroscopy and Radiative Transfer},
volume = {347},
pages = {109596},
year = {2025},
issn = {0022-4073},
doi = {https://doi.org/10.1016/j.jqsrt.2025.109596},
url = {https://www.sciencedirect.com/science/article/pii/S0022407325002584},
author = {P. Wcisło and N. Stolarczyk and M. Słowiński and H. Jóźwiak and D. Lisak and R. Ciuryło and A. Cygan and F. Schreier and C.D. Boone and A. Castrillo and L. Gianfrani and Y. Tan and S.-M. Hu and E.M. Adkins and J.T. Hodges and H. Tran and H.N. Ngo and J.-M. Hartmann and S. Beguier and A. Campargue and R.J. Hargreaves and L.S. Rothman and I.E. Gordon}
}

@article{Kauten2014,
  author  = {Kauten, T. and Pressl, B. and Kaufmann, T. and Weihs, G.},
  title   = {Measurement and modeling of the nonlinearity of photovoltaic and {Geiger}-mode photodiodes},
  journal = {Rev. Sci. Instrum.},
  volume  = {85},
  pages   = {063102},
  year    = {2014},
  doi     = {10.1063/1.4879820}
}

@book{Lakowicz2006, 
author={Lakowicz, Joseph R.}, 
title={Principles of Fluorescence Spectroscopy}, 
edition={3rd}, 
publisher={Springer}, 
address={New York}, 
year={2006}
}

@inproceedings{Gong2007,
  author    = {Gong, Yuan and Li, Bincheng},
  title     = {Effect of instrumental response time in exponential-decay based cavity ring-down techniques for high reflectivity measurement},
  booktitle = {Laser-Induced Damage in Optical Materials: 2007},
  editor    = {Exarhos, Gregory J. and Guenther, Arthur H. and Lewis, Keith L. and Ristau, Detlev and Soileau, M. J. and Stolz, Christopher J.},
  series    = {Proc. SPIE},
  volume    = {6720},
  pages     = {67201E},
  year      = {2007},
  doi       = {10.1117/12.758948}
}

\bigskip
\noindent\textbf{Acknowledgments:} For the purpose of Open Access, the authors have applied a CC-BY public copyright license to any author accepted manuscript (AAM) version arising from this submission. The research is part of the program of the National Laboratory FAMO in Toru\'n, Poland.
\textbf{Funding:} The research was supported by the National Science Centre, Poland, Projects No. 2020/39/B/ST2/00719 (R.C. and A.C.), 2021/42/E/ST2/00152 (S.W.), and 2023/51/B/ST2/00427 (D.L.), and by the state budget of Poland, allocated by the Minister of Education and Science under the ``Polska Metrologia II'' program, project no. PM-II/SP/0011/2024/02, subsidy amount 988\,900 PLN. We acknowledge partial funding support from the MetCTG project. The project (24GRD06 MetCTG) has received funding from the European Partnership on Metrology, co-financed from the European Union’s Horizon Europe Research and Innovation Programme and by the Participating States.
\textbf{Author contributions:} Conceptualization: A.C., D.L.; Methodology: A.C.; Software: R.C., A.C., M.G., J.Ł.; Formal analysis: R.C., A.C.; Investigation: A.C., D.L., S.W., J.Ł.; Validation: A.C., R.C.; Visualization: A.C., R.C.; Funding acquisition: A.C., D.L., S.W.; Project administration: A.C., D.L.; Supervision: A.C.; Writing---original draft: A.C.; Writing---review and editing: A.C., D.L., S.W., R.C..
\textbf{Competing interests:} The authors declare that they have no competing interests.
\textbf{Data and materials availability:} All data needed to evaluate the conclusions in the paper are present in the paper.

\end{document}